# The Fireball of November 24, 1970, as the Most Probable Source of the Ischgl Meteorite

Maria Gritsevich[1,2,3], Jarmo Moilanen[1,2], Jaakko Visuri[2], Matthias M. M. Meier[4,5], Colin Maden[4], Jürgen Oberst[6], Dieter Heinlein[7], Joachim Flohrer[8], Alberto J. Castro-Tirado[9,10], Jorge Delgado-García[11], Christian Koeberl[12], Ludovic Ferrière[13], Franz Brandstätter[13], Pavel P. Povinec[14], Ivan Sýkora[14], Florian Schweidler[15]

1. Faculty of Science, University of Helsinki, Gustaf Hallströmin katu 2, FI-00014 Helsinki, Finland
2. Finnish Fireball Network, Ursa Astronomical Association, Kopernikuksentie 1, FI-00130 Helsinki, Finland
3. Institute of Physics and Technology, Ural Federal University, Mira str. 19, 620002 Ekaterinburg
4. Institute of Geochemistry and Petrology, ETH Zurich, Clausiusstrasse 25, CH-8092 Zurich, Switzerland
5. Naturmuseum St.Gallen, Rorschacher Strasse 263, CH-9016 St.Gallen, Switzerland
6. Institute of Planetary Research, Technical University Berlin, Strasse des 17. Juni 135, D-10623 Berlin, Germany
7. German Fireball Network, Lilienstrasse 3, D-86156 Augsburg, Germany
8. Institute of Planetary Research, German Aerospace Center (DLR), Rutherfordstrasse 2, D-12489 Berlin, Germany
9. Instituto de Astrofísica de Andalucía (IAA-CSIC), Glorieta de la Astronomía s/n, E-18008, Granada, Spain
10. Departamento de Ingeniería de Sistemas y Automática, Escuela de Ingenierías, Universidad de Málaga, c/ Dr. Ortiz Ramos sn, E-29071, Málaga, Spain
11. Departamento de Ingeniería Cartográfica, Geodésica y Fotogrametría, Universidad de Jaén, Campus Las Lagunillas s/n Edificio A3. E-23071 Jaén, Spain
12. Department of Lithospheric Research, University of Vienna, Josef-Holaubek-Platz 2, A-1090 Vienna, Austria
13. Natural History Museum Vienna, Burgring 7, A-1010 Vienna, Austria
14. Faculty of Mathematics, Physics and Informatics, Comenius University, SK-84248 Bratislava, Slovakia
15. School of Mechanical Engineering, Marienstrasse 1, D-84028 Landshut, Germany

**Abstract**

The discovery of the Ischgl meteorite unfolded in a captivating manner. In June 1976, a pristine meteorite stone weighing approximately 1 kg, fully covered with a fresh black fusion crust, was collected on a mountain road in the high-altitude Alpine environment. The recovery took place while clearing the remnants of a snow avalanche, 2 km northwest of Ischgl in Austria. Subsequent to its retrieval, the specimen remained in the finder's private residence without undergoing any scientific examination or identification until 2008, when it was brought to the University of Innsbruck. The sample was classified as a well-preserved LL6 chondrite, with a W0 weathering grade, implying a relatively short time between the meteorite fall and its retrieval. To investigate the potential connection between the Ischgl meteorite and a recorded fireball event, we have reviewed all documented fireballs ever photographed by German fireball camera stations. This examination led us to identify the fireball EN241170 observed in Germany by ten different European Network stations on the night of November 23/24, 1970, as the most likely candidate. We employed state-of-the-art techniques to reconstruct the fireball's trajectory, and to reproduce both its luminous and dark flight phases in detail. We find that the determined strewn field and the generated heat map closely align with the recovery location of the Ischgl meteorite. Furthermore, the measured radionuclide data reported here indicate that the pre-atmospheric size of the Ischgl meteoroid is consistent with the mass estimate inferred from our deceleration analysis along the trajectory. Our findings strongly support the conclusion that the Ischgl meteorite originated from the EN241170 fireball, effectively establishing it as a confirmed meteorite fall. This discovery enables to determine, along with the physical properties, also the heliocentric orbit and cosmic history of the Ischgl meteorite.

## 1. Introduction

In June 1976, a fortuitous discovery was made on a mountain road about 2 km northwest of the Tyrolean town of Ischgl in Austria. A single stone, fully covered with a fresh fusion crust, was found by Josef Pfefferle, a professional forest ranger, while clearing the remnants of a snow avalanche. According to the finder, the stone had apparently fallen out of the snow and was lying in the middle of the road (Meteoritical Bulletin, no. 101, Brandstätter et al. 2013). Weighing approximately 1 kg, the peculiar appearance of the stone, with its dull black crust, caught Mr. Pfefferle's attention. However, it was not until 2008 that Mr. Pfefferle showed the specimen to scientists, who confirmed its meteoritic origin.

The recovery site is located at coordinates $\varphi = 47° 01' 34.8" = 47.02633°$ N, $\lambda = 10° 16' 24.0" = 10.27333°$ E, with an altitude of 2000 meters (WGS84 datum; determined using SRTM Worldwide Elevation Data at a resolution of 1 arc-second). As described by Brandstätter et al. (2013), the fist-sized stone had apparently been displaced from a higher altitude by a snow avalanche. Although Mr. Pfefferle recognized its unusual appearance, he initially disregarded the possibility of it being a meteorite, placed the rock into a box, and kept it at his home for more than thirty years. In 2007, Austrian news media reported on a court case involving a meteorite (the third recovered fragment of the Neuschwanstein meteorite) found near Reutte, in Tyrol, Austria. These reports reminded Mr. Pfefferle of his find in the year 1976. In 2008, he took the sample to the University of Innsbruck, where the meteoritic nature of the rock was finally established.

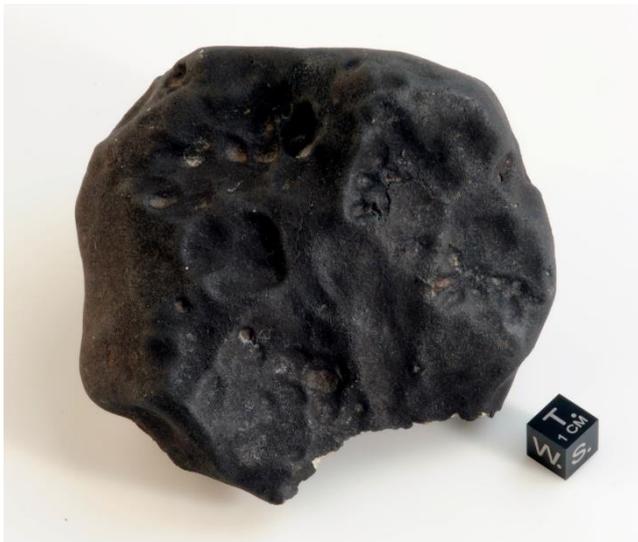
a

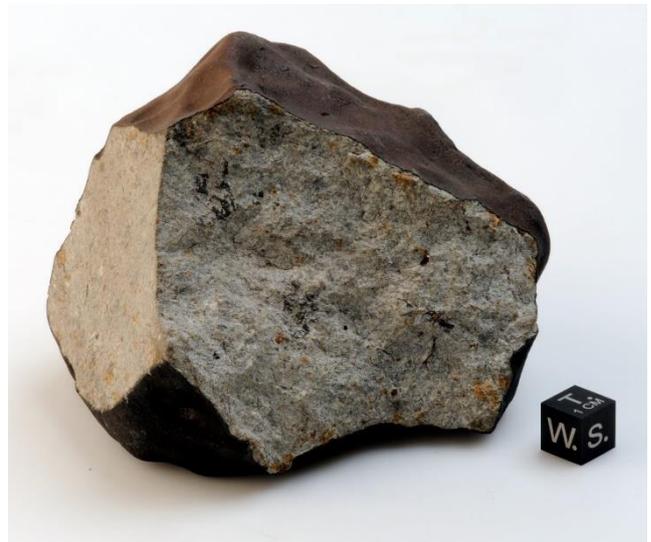
b

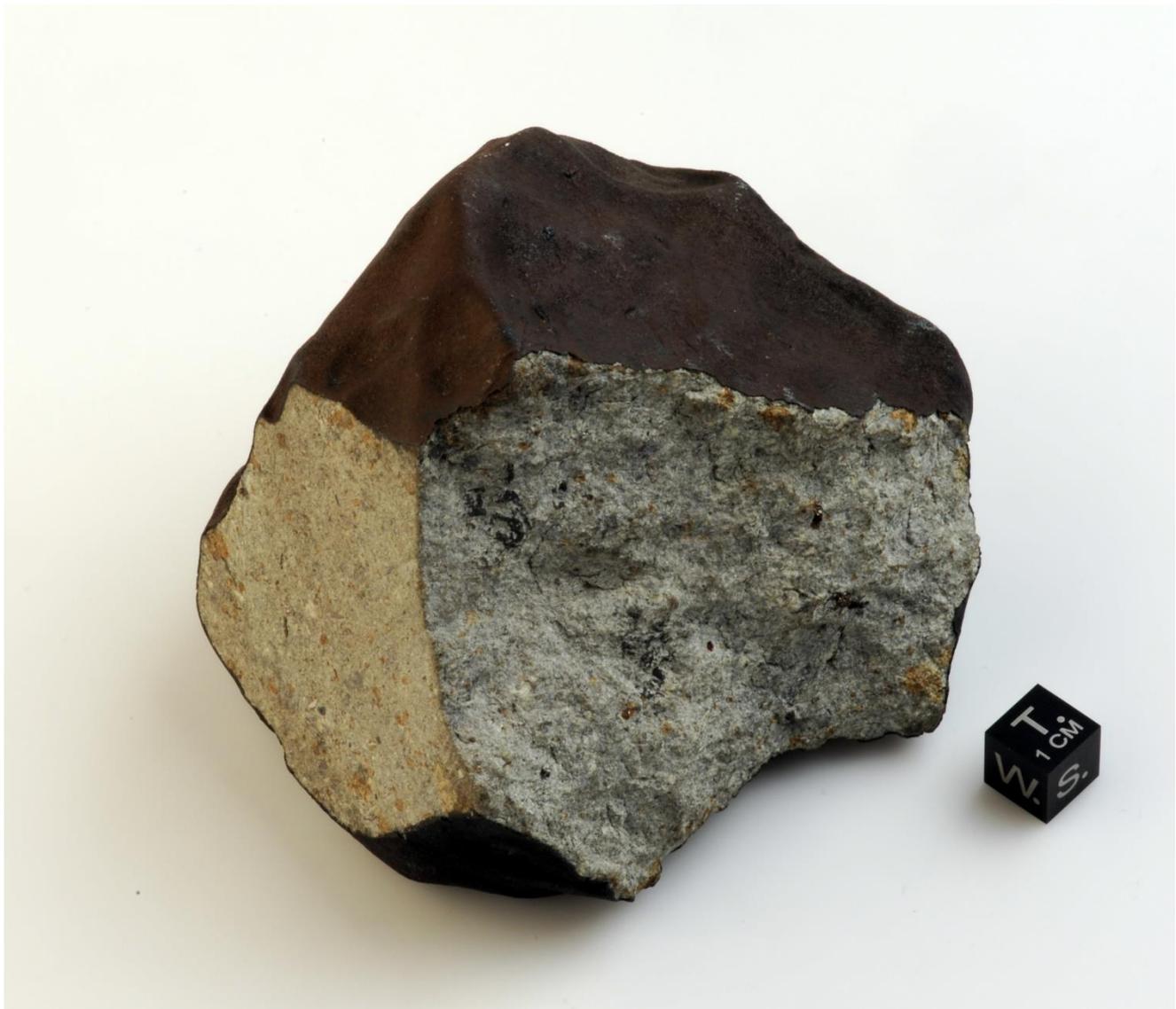

c

Figure 1. The Ischgl meteorite classified as a type LL6 (S3, W0) ordinary chondrite. (a) Fusion crusted exterior of the main mass, weighing 708.1 g, and exhibiting numerous well-developed regmaglypts. (b) and (c) Views showing its light-grey colored interior. A broken face (~8 × 5 cm) exhibits a light-gray breccia with few metallic grains. A few patchy brown oxidized areas are visible on the cut and broken surface. Sample NHMW-N9269 from the Natural History Museum Vienna's collection.

In 2011, following an initiative of two of the authors of this study (CK, FB), the Ischgl meteorite was acquired by the Natural History Museum in Vienna. Since November 2012, it has been displayed in the museum's Meteorite hall, in a special showcase with all meteorites recovered so far in Austria. An estimated mass of about 200–300 g is missing from the original main mass, as parts of it were chipped off early on using a hammer. Incidentally, a small fragment was broken off the main mass by a friend of the finder and allegedly presented as a donation to the Tiroler Landesmuseum Ferdinandeum in Innsbruck already in 1989. Although the meteoritic nature of the sample was officially confirmed only in 2008, this donation to the Innsbruck Museum was properly declared as a meteorite (Anonymous 1990). Unfortunately, this fragment of the Ischgl meteorite is no longer available at the Ferdinandeum Museum, and there is also no record of its mass, although doubts have been raised if the sample was actually handed over in 1989.

A small piece was cut off later at the University of Innsbruck (Brandstätter et al. 2013). The remaining main mass of the meteorite, with a weight of 708.1 g, measures approximately 9 × 8 × 6 cm. A 3D shape model of the specimen is provided in the Supplementary Material. Ischgl was classified as an ordinary chondrite of type LL6, S3, W0, implying excellent preservation conditions and a relatively short period of time between the meteorite fall and the recovery (Garvie et al. 2013). The bulk density of the specimen was determined as 3.31 g/cm$^3$ using the method of measuring bulk volume with glass beads.

The exterior of the meteorite is completely covered with fusion crust (except for the parts that have been chipped and cut off) and exhibits well developed regmaglypts (Fig. 1a). The fusion crust appears extremely fresh and has a dull black color. The exposed interior of the meteorite shows a light-grey colored rock without any distinct clasts. The cut surface exhibits a chondritic texture with poorly defined chondrule-like outlines and finely dispersed small metal grains (Fig. 1b). It is clear that the meteorite fall must have happened relatively shortly prior to the recovery of the specimen.

The original goal of this study was to revisit all notable fireball events ever photographed by the German EN camera stations over Central Europe with the support of today's knowledge and to reanalyze if and where they would be expected to produce meteorites. In the following sections we detail how an application of the state-of-the-art data processing techniques allowed us to robustly identify the missing link between the bright fireball EN241170 (Mount Riffler) and the Ischgl meteorite. Using the high-resolution photogrammetric scans of the original films for measuring the fireball flight directions and velocity distribution, we determined the best fitting solution for the atmospheric trajectory. This solution and historical atmospheric data for the time and location of the fireball event were used as inputs to run the dark flight Monte Carlo (DFMC) simulations (Moilanen et al. 2021; Gritsevich et al. 2021). The simulated strewn field, as well as a heat map, indicate the probable distribution of masses and landing positions of the meteorite fragments resulting from the November 24, 1970 fireball. Additionally, we obtained radionuclide data, which also allow an intercomparison of pre-atmospheric size estimates, further strengthening the connection between the fireball EN241170 and the Ischgl meteorite.

## 2. Observational data

In order to gather extensive data on orbits of Solar System small bodies and fireball trajectories and their behaviors, photographic all-sky camera networks have been established worldwide since the mid-20th century (Whipple 1938; Ceplecha and Rajchl 1965; Halliday et al. 1978; Trigo-Rodríguez et al. 2006; Bland et al. 2012; Howie et al. 2017; Colas et al. 2020; Devillepoix et al. 2020). Subsequently, a possibility of meteor detection has been discussed or even integrated also within the broader-scoped surveys on Earth and beyond including the Vera C. Rubin Observatory, previously referred to as the Large Synoptic Survey Telescope (LSST), Mini-EUSO telescope on board the ISS, High-Power Large-Aperture (HPLA) radars such as the European Incoherent Scatter facility EISCAT, the Global Network of Robotic Astronomical Observatories BOOTES, and EnVision mission to Venus developed by the European Space Agency (Pellinen-Wannberg et al. 1998, Vinković et al., 2016, Räbinä et al. 2016, Bektešević et al. 2018, Kero et al. 2019, Vinković and Gritsevich 2020, Coleman et al. 2023, Castro-Tirado 2023, Christou and Gritsevich, 2024). Of particular interest are relatively rare meteorite-producing events (Wetherill and ReVelle 1981; Gritsevich 2008a; Brown et al. 2013; Sansom et al. 2019; Boaca et al. 2022; Peña-Asensio et al. 2023), as their identification and subsequent analysis may allow rapid recovery of unique extraterrestrial samples in unweathered conditions whilst also providing clues to their dynamic origin (Dmitriev et al. 2015; Trigo-Rodríguez et al. 2015; Jansen-Sturgeon et al. 2019; Peña-Asensio et al. 2021; Kyrylenko et al. 2023). Well-preserved meteorites with known orbits are central to our understanding of the conditions and processes that defined the history of the Solar System.

The European Fireball Network (EN) began its operation in 1963 using all-sky cameras in (what was then) Czechoslovakia under the direction of Zdeněk Ceplecha at the Ondřejov Observatory of the Astronomical Institute of the Czechoslovak Academy of Sciences (Ceplecha, 1977). Three years later, a number of similar stations for the systematic fireball registration were deployed in Germany (Zähringer, 1969a; 1969b; Oberst et al., 1998). These 25 unguided semiautomatic all-sky survey cameras were set up mainly in Southern Germany (Fig. 2, see also Supplementary Material). Recognizing their profound importance in achieving our goal of identifying and reducing the data on fireballs that could potentially match the Ischgl case, we conducted a thorough analysis of their records and revisited their operational principles. As the German network's operation came to a close in 2022, and considering the inevitable decline in data accessibility over time and with changing generations, we provide a concise summary of these data within the present study.

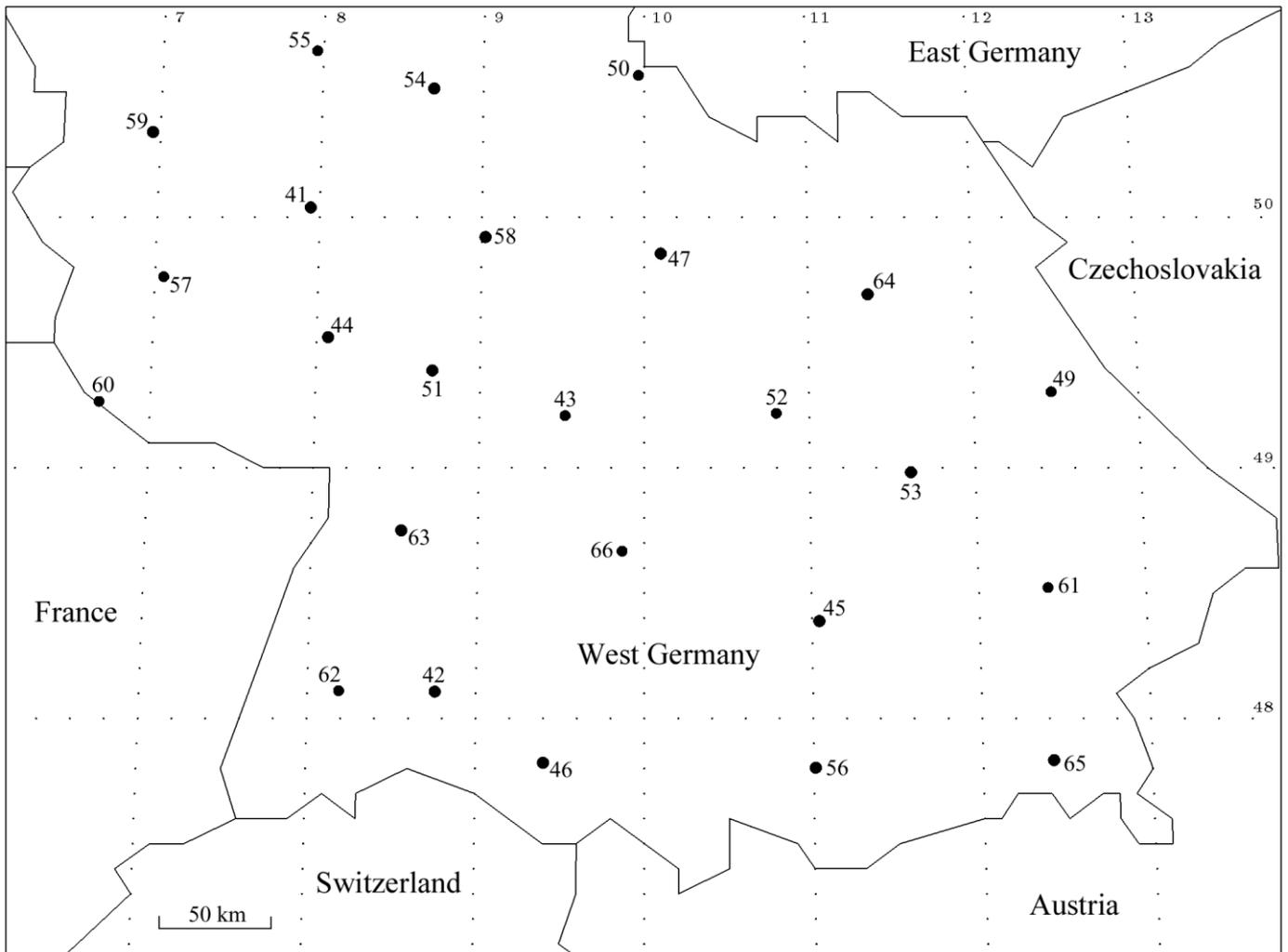

Figure 2. A map displaying the locations of the 25 meteor cameras operated by the MPIK, predominantly concentrated in Southern Germany, during the year 1970. The German part of the European Fireball Network was comprised of the following stations: 41 Stephanshausen, 42 Klippeneck, 43 Öhringen, 44 Wattenheim, 45 Dasing, 46 Glashütten, 47 Seligenstadt, 49 Neukirchen, 50 Eckweisbach, 51 Heidelberg, 52 Mitteleschenbach, 53 Zell, 54 Leihgestern, 55 Marienberg, 56 Hohenpeißenberg, 57 Deuselbach, 58 Schaafheim, 59 Nürburg, 60 Berus, 61 Gerzen, 62 Schönwald, 63 Wildbad, 64 Obertrubach, 65 Bernau, and 66 Stötten.

In the beginning, the German segment of the EN was overseen by the Max-Planck-Institut für Kernphysik (MPIK) in Heidelberg. Subsequently, its management transitioned to the Institute of Planetary Research (Berlin-Adlershof) of the German Aerospace Center (Deutsches Zentrum für Luft- und Raumfahrt, DLR) and to the Technical University of Berlin. Initially, the instrumental setup of the German cameras closely resembled the Czechoslovak system. The MPIK stations were equipped with Leica MD cameras featuring 50 mm Zeiss objectives and using regular 24 x 36 mm greyscale film. These cameras captured images of an all-sky parabolic mirror with a 36 cm diameter (Figs. 3 and 4). To capture fast-moving objects accurately, all stations were equipped with 12.5 Hz rotating shutters, which resulted in interrupted trails. This allowed for determining the duration of the fireballs and the angular velocities of the meteoroids. Each night, a single exposure was taken, and the timing information to open and close the aperture correctly was obtained from analog clocks. Local operators programmed time switches (see Fig. 5, left image) according to a specified schedule (Oberst et al., 1998).

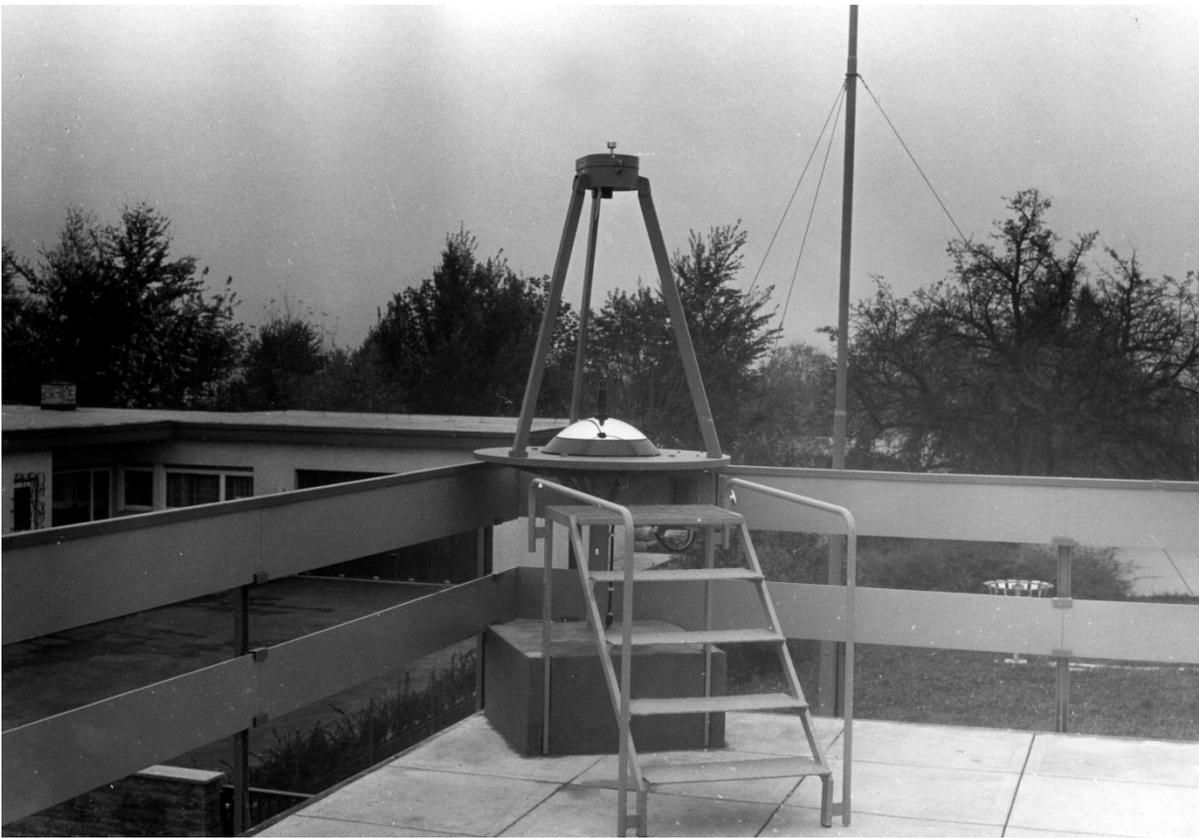

Figure 3. EN 43 Öhringen station was set up in November 1966 on the flat roof of the Öhringen weather service station. This all-sky camera remained in continuous operation at its original location until July 2021. Picture credit: Max-Planck-Institut für Kernphysik, Heidelberg.

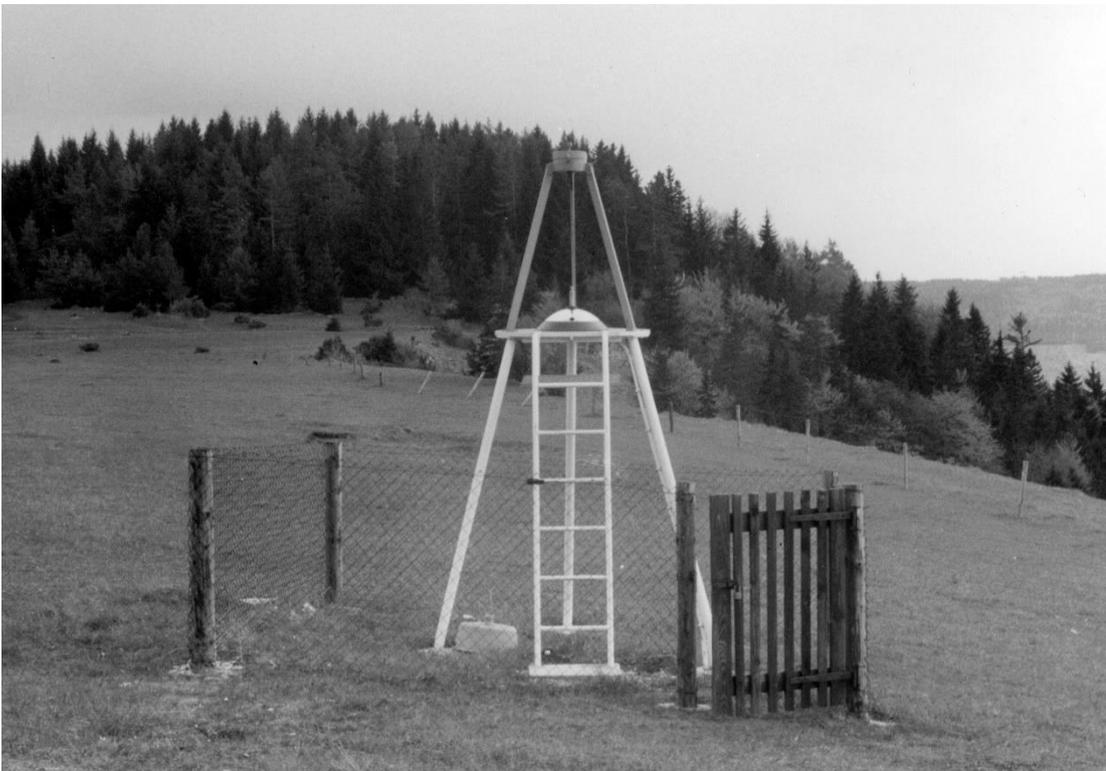

Figure 4. EN 61 Gerzen tripod station was erected in November 1967 and operated until 1988. During the process of rearranging and expanding the German part of the EN, the station was relocated to a different location. Picture credit: Max-Planck-Institut für Kernphysik, Heidelberg.

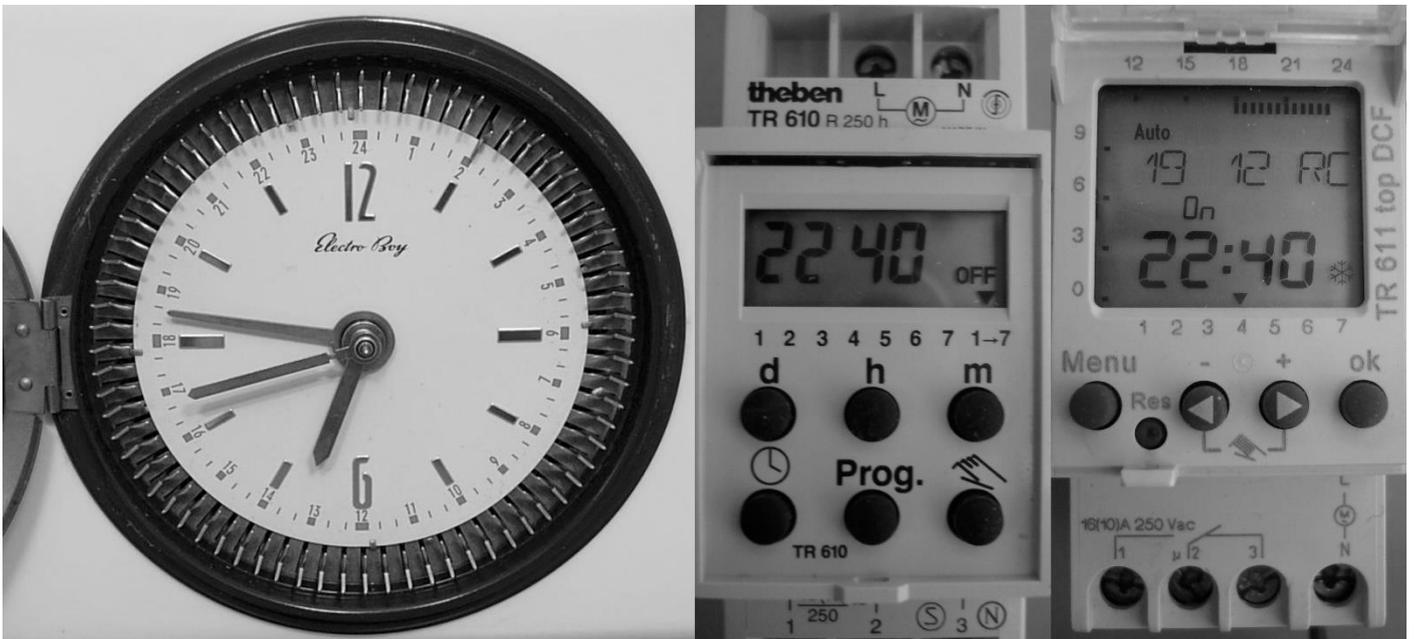

Figure 5. Analog switch clocks (left) were used in the German all-sky stations from 1966 until 1988, while the MPIK was in charge of the fireball stations. During the network's reorganization in the 90s, digital time switches (center) replaced them. Subsequently, radio-controlled autotimers (right) were responsible for triggering the long time exposures of the meteor cameras.

Although the German part of the European Network was managed by the MPIK, the reduction and analysis of the meteor data of interest were conducted by Zdeněk Ceplecha and his team at Ondřejov. Not all of the recorded data were processed during this analysis. The results were then published jointly (e.g., Zähringer, 1969a; Ceplecha et al., 1973; 1976). Over the years, under the management of the DLR Institute of Planetary Research, the German part of the EN underwent various modifications, including changes in geographic positions, spatial density, and equipment for fireball detection stations. In the end, 15 of the original 25 mirror cameras remained operational until the spring of 2022.

### 3. Meteorite-producing events

Because the German EN fireball camera stations have been in operation since 1966, we took the opportunity to revisit all potential meteorite-dropping events. Despite the fact that the Ischgl meteorite could only be matched with a fireball event occurring prior to its recovery, we recognize the added value in summarizing the outcomes of over 50 years of observations within the area of coverage, for accessing the likelihood from a statistical perspective.

During the continuous operation of the German EN stations over many decades, more than two thousand fireballs were registered and documented. The vast majority of the meteoroids fully disintegrated in the atmosphere. Only in a dozen cases, fireball events classified as a promising meteorite fall have been instrumentally recorded. These fireball events (with estimated terminal masses over 0.5 kilograms) that were simultaneously photographed by at least two German EN cameras and may have been of interest for the current study are listed in Table 1. A summary of their earlier predicted fall locations is shown in Fig. 6 (see Oberst et al. 1998, 2004; Spurný 2002, 2003, 2010, 2011; Heinlein 2013a, 2013b; Gritsevich and Stulov 2008; Heinlein and Spurný 2012; Bischoff et al. 2017, 2019; Borovička 2021; Moilanen et al. 2021 for more insights).

Table 1. Presumed meteorite-dropping fireballs, simultaneously photographed by German EN camera stations over South Central Europe with estimated terminal masses exceeding 0.5 kilograms (Oberst et al., 1998, 2004; Heinlein, 1992, 2013a, 2013b, 2014a, 2014b, 2014c, 2016, 2019, 2020; Heinlein and Spurný, 2012; Spurný et al. 2002, 2003; Braun 2015).

| Date of meteor event yyyy-mm-dd | Time (UT) | Number of stations | Name | Terminal mass (kg) | Calc. impact location Lat. (N) | Long. (E) |
|---|---|---|---|---|---|---|
| 1969-04-10 | 21:45 | 4  | Otterskirchen | 5.0  | 48.650° | 13.333° |
| 1970-11-24 | 01:47 | 10 | Mount Riffler | 0.9  | 47.133° | 10.350° |
| 1974-08-30 | 01:25 | 6  | Leutkirch | 9.6  | 47.850° | 09.900° |
| 1977-06-01 | 21:46 | 6  | Freising | 0.7  | 48.467° | 11.650° |
| 1977-06-12 | 23:02 | 2  | The Alps | 30.0 | 46.100° | 06.483° |
| 1993-02-22 | 22:13 | 8  | Meuse | 2.7  | 49.417° | 04.800° |
| 2002-04-06 | 20:20 | 8  | Neuschwanstein[1] | 6.9 | 47.530° | 10.805° |
| 2012-02-21 | 21:00 | 3  | Stanzach | 0.5  | 47.379° | 10.476° |
| 2015-03-15 | 19:44 | 8  | Airolo | 9.0  | 46.519° | 08.644° |
| 2015-06-02 | 21:51 | 2  | Neudes | 0.5  | 50.100° | 11.991° |
| 2016-03-06 | 21:37 | 2  | Stubenberg[2] | 10.0 | 48.295° | 13.117° |
| 2018-07-10 | 21:30 | 3  | Renchen[3] | 4.0  | 48.588° | 08.000° |

[1] Three fragments of this meteorite fall (1750, 1625, and 2840 g) were recovered after systematic search campaigns in the calculated impact area.
[2] Six fragments of this meteorite fall (48, 8, 19, 42, 1320, and 36 g) were recovered after systematic search campaigns in the calculated impact area.
[3] Six fragments of this meteorite fall (12, 955, 21, 5, 6, and 228 g) were recovered after systematic search campaigns in the calculated impact area.

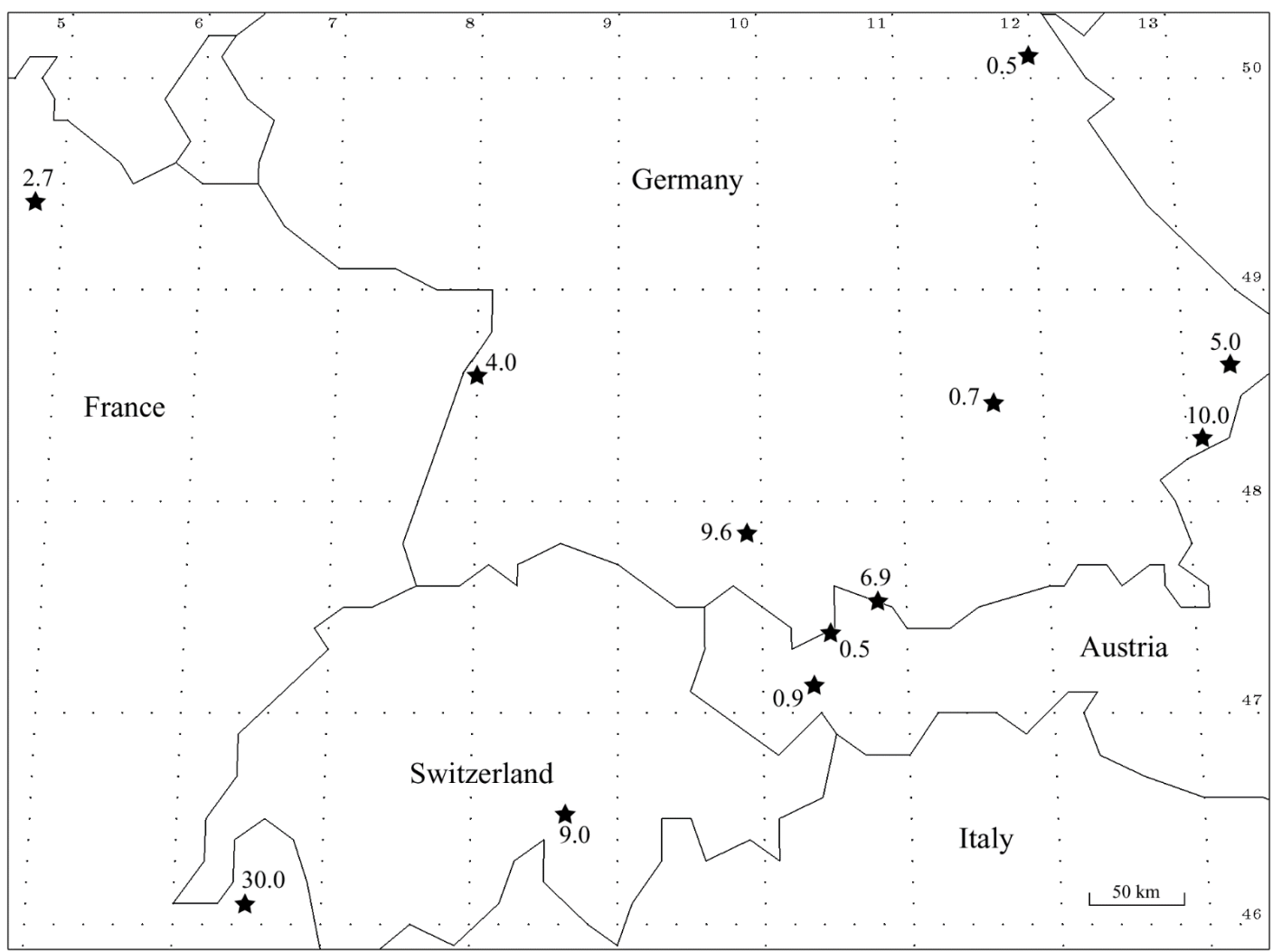

Figure 6. All meteorite-dropping fireballs with predicted remaining mass of 0.5 kg or more that have been simultaneously registered by the German all-sky cameras to date are shown over the indicated area. The calculated impact sites are marked by stars, and the estimated terminal masses are indicated in kilograms.

## 4. The November 24, 1970 fireball

During the night of November 23/24, 1970, a bright fireball EN241170 was photographed by multiple EN stations. The fireball occurred at $01^h 47^m$ UT = $02^h 47^m$ MEZ. The passage time of the meteor was reported by visual observers with a precision of $\pm 1^m$ (Ceplecha, 1977; Oberst et al., 1998). The preliminary results of the meteorite-dropping event EN241170 were briefly reported shortly after its occurrence (Ceplecha et al., 1971). In this early publication, only limited data regarding the event were provided, including a terminal mass of 2 kg, a possible fall location near Innsbruck, Austria, and an aphelion distance of the heliocentric orbit (Q) of 1.51 au.

In a review report on bright fireballs registered by Czech and German EN cameras, Ceplecha (1977) further described the event as a bright fireball of magnitude -15, suggesting a possible meteorite fall with a terminal mass of 0.9 kg and specified its orbital parameters. The author concluded that the EN241170 fireball was of "type I" according to the introduced back then classification, indicating that the meteoroid likely originated from an asteroid and its fragments most likely have survived the atmospheric entry, reaching the ground.

Table 1 in (Ceplecha, 1977) lists the value of $\cos(z_R) = \sin\gamma$ of 0.310 for the Mt. Riffler fireball. The predicted fall location was determined by computing the dark flight trajectory after the terminal point. This involved using the calculated terminal values of velocity, deceleration, and air density, along with considerations of the wind field (Ceplecha, 1977). The calculations were based on solving the ballistic flight equations for the idealized assumption of a single fragment of a spherical shape, and not taking into account additional forces able to alter the meteoroid's trajectory.

The geographical coordinates of the German EN stations, including latitude, longitude, and elevation, were sourced from topographic maps available at the time of the study. The coordinates refer to the previously used Potsdam datum, and, therefore, all results likewise refer to this geodetic datum. The meteorite fall location predicted by Ceplecha (1977) is $\varphi$ = 47° 08' 26" N, $\lambda$ = 10° 20' 44" E (Potsdam datum) and corresponds to $\varphi$ = 47° 08' 23.0" = 47.13972° N, $\lambda$ = 10° 20' 39.7" = 10.34436° E (WGS84 datum). This fall site is located near the town of Pettneu am Arlberg, in Tyrol (Austria), and approximately 3 km from the summit of the mountain known as "Hoher Riffler" (altitude 3168 meters). Therefore, the possible meteorite fall EN241170 was given the provisional name "Mount Riffler" (Table 1).

Due to the shallow entry of the fireball, its lengthy atmospheric trajectory, and the large associated area of the probable meteorite fall, as well as the challenging mountainous terrain around Mount Riffler, no systematic field search campaign was ever undertaken for the EN241170 case.

## 5. Photographic data reduction

For this study, the photographic fireball EN241170 data were thoroughly reanalyzed using modern processing techniques. In the meteor archive of the MPIK, which is now hosted by the DLR in Augsburg, 10 original negatives of this extraordinary meteor event still exist (Fig. 7). These 10 photographs were taken at the all-sky camera sites 61 Gerzen (Fig. 7a), 66 Stötten (Fig. 7b), 52 Mitteleschenbach (Fig. 7c), 43 Öhringen, 62 Schönwald (Fig. 7d), 49 Neukirchen, 50 Eckweisbach, 57 Deuselbach, 60 Berus, and 55 Marienberg. Six of these negatives were digitized with high quality using a Vexcel UltraScan 5000 photogrammetric scanner, which has a resolution of 5080 ppi, corresponding to a pixel size of 5 micrometers. The scanner employs transmitted light for digitizing negatives and was calibrated beforehand to ensure the accuracy and geometric fidelity of the resulting digital images. The images were stored in TIFF format without any compression, ensuring the preservation of their original quality.

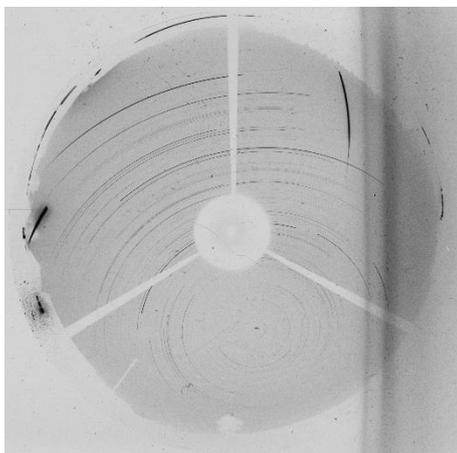
Figure 7a. The all-sky image captured by EN camera 61 Gerzen.

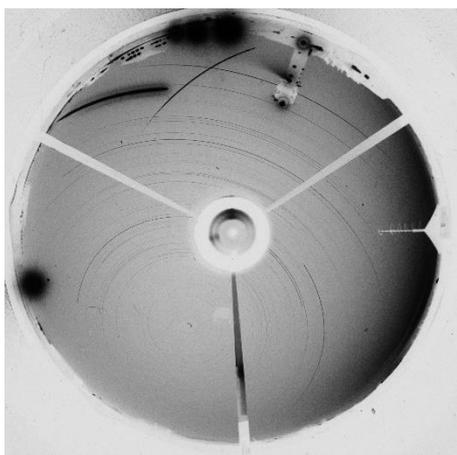
Figure 7b. The all-sky image captured by EN camera 66 Stötten.

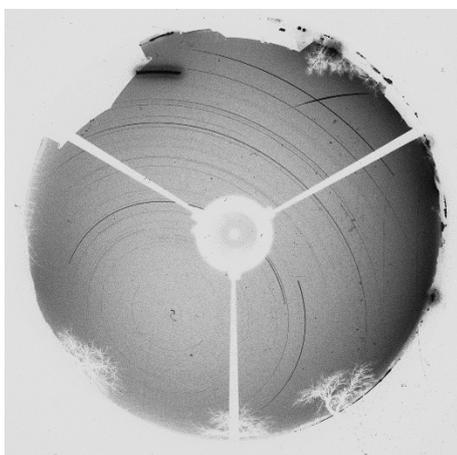
Figure 7c. The all-sky image captured by EN camera 52 Mitteleschenbach.

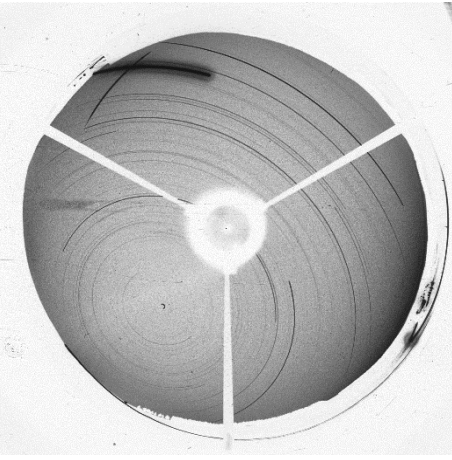

Figure 7d. The all-sky image captured by EN camera 62 Schönwald.

Figure 7. The all-sky images of the EN241170 fireball captured during the night of November 23/24 and used throughout this study. These images were acquired by the stations: No. 61 Gerzen (a), 66 Stötten (b), 52 Mitteleschenbach (c), and 62 Schönwald (d).

Table 2. Coordinates of the ten EN stations that recorded the EN241170 fireball. The original coordinates, initially in the Potsdam datum, have been converted to GPS coordinates referencing the WGS84 datum. The complete list of station coordinates can be found in the supplementary material.

| EN | Location | Latitude (N) | Longitude (E) | Elevation |
|---|---|---|---|---|
| 43 | Öhringen | 49.20672° | 09.51806° | 280 m |
| 49 | Neukirchen | 49.30203° | 12.46847° | 490 m |
| 50 | Eckweisbach | 50.56436° | 09.96169° | 525 m |
| 52 | Mitteleschenbach | 49.21589° | 10.79772° | 411 m |
| 55 | Marienberg | 50.66211° | 07.95769° | 545 m |
| 57 | Deuselbach | 49.76108° | 07.05283° | 480 m |
| 60 | Berus | 49.26392° | 06.68942° | 365 m |
| 61 | Gerzen | 48.52117° | 12.41183° | 460 m |
| 62 | Schönwald | 48.10961° | 08.18478° | 990 m |
| 66 | Stötten | 48.66622° | 09.86525° | 732 m |

To accurately recover the trajectory parameters of the fireball captured by the camera stations, it is essential to establish the geometric relationship between the reference system of each camera station and the horizon coordinate system (Ceplecha, 1987; Nagel, 2011; Trigo-Rodríguez et al., 2015; Lyytinen and Gritsevich, 2016a). This involves determining the camera positions and orientations. Additionally, image distortions caused by the curvature of the parabolic mirrors must be considered. To obtain the camera coordinates in the current coordinate system, the previously reported Potsdam datum of the EN stations were converted to the WGS84 datum, which is typically realized through GNSS measurements today (Table 2).

As the actual orientations of the cameras at the time of the event are only roughly known, calibration was performed using background stars that appear in the photographs. Due to the long image exposures, stars are smeared and form long arcs around the north celestial pole. Thus, knowing the beginning and end times of the photograph exposures is crucial for recovering star positions on the photographs at any given time. The exposure times were specified by the MPIK, and records of these exposure times are still available today. For the analog switch clocks used in the German all-sky stations (which have since been replaced in the current camera stations, see Fig. 5), exposure times were typically off by several minutes. In the case of EN station 61 Gerzen, exposure times were mistakenly offset by several hours due to a human operator error.

The images captured by EN camera stations 52, 61, 62, and 66 were calibrated to determine the luminous flight parameters and reconstruct the trajectory using the FireOwl, our latest software developed for processing data acquired by the Finnish Fireball Network (Visuri and Gritsevich, 2021; Kyrylenko et al., 2023; Peña-Asensio et al. 2024). As explained in the following section, data from EN camera station 61 were utilized twice. Initially, it was employed to establish the triangulated fireball trajectory with precise astrometry, focusing on the shorter path of the fireball (Fig. 8). Subsequently, velocity values were derived using the longer path of the fireball, whereas the astrometry for the end part of the flight was not as accurate.

The setup of the EN camera stations followed the description provided by Oberst et al. (1998). The all-sky cameras employed convex mirrors, which produce mirror images. For calibration purposes, the images were reduced accordingly, and their orientation was corrected by rotating them by 90 degrees multiple times until Polaris was located in the left bottom quarter sector of every image.

Equatorial coordinates of stars were adopted from epoch J2000.0 and underwent precession correction backward in time to correspond to the event's timing. Manual clicks were made to identify the positions of stars and Saturn at the start or end of the star trails, along with their corresponding times.

The radial distortion of objects in the image was described using Equation (1), which accounts for further adjustments to ensure the accuracy and reliability of the trajectory data obtained from the calibration process:

$$R = (1-t)*\text{atan}(r/f) + t*(r/f) + k_5*(r/f)^5 + k_7*(r/f)^7 \qquad (1)$$

In Eq. (1) R represents the real angle in radians between the image optical axis and the object, while r corresponds to the observed radial distance from the image center in pixels. The parameter t refers to the barrel parameter or "radial distortion parameter", and f denotes the focal length in pixels. This correction equation, which effectively accounts for radial distortion, was previously introduced in the work of Lyytinen and Gritsevich (2016a) and was previously used for the trajectory retrieval leading to the successful meteorite recovery, such as in the cases of Annama and Ozerki (Trigo-Rodríguez et al. 2015; Lyytinen and Gritsevich 2016b; Maksimova et al. 2021).

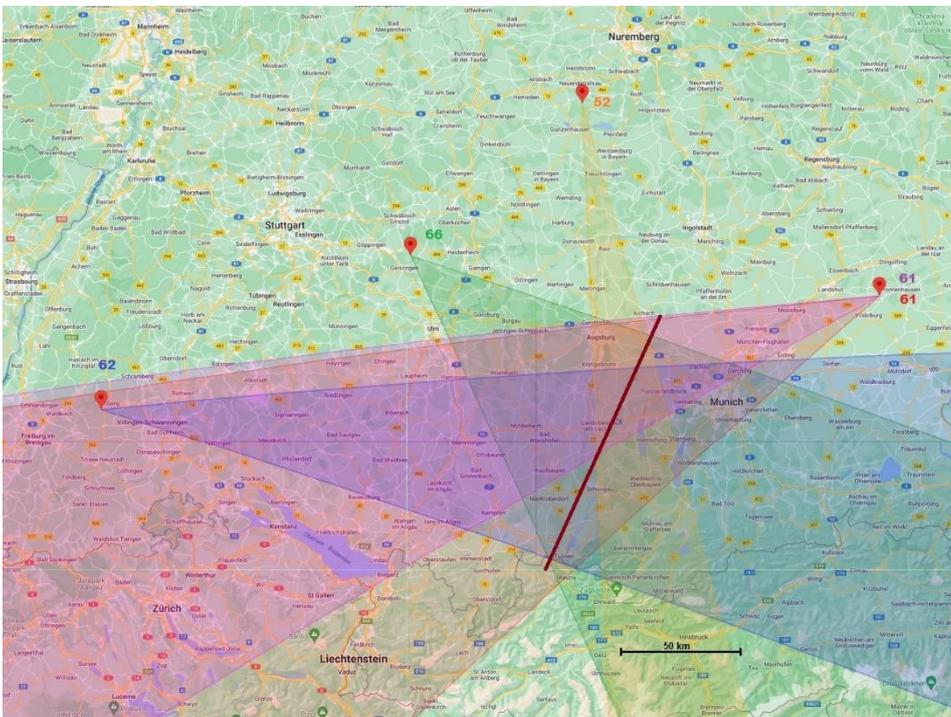

Figure 8. The fireball trajectory of EN241170, as observed from the selected stations (Table 2), along with the station orientations. The two shaded regions originating from the camera station 61 illustrate the extent to which trajectory data from this station were utilized for the initial and secondary analyses. The narrower area corresponds to triangulation, while the second area represents velocity distribution calculations.

To achieve further improvement in the precision of the trajectory reconstruction, additional correction terms of the 5th and 7th degree were incorporated. By introducing these extra correction terms, the calibration process becomes more robust and accurate, allowing for better refinement of the trajectory data obtained from the image analysis. This enhanced calibration approach was particularly important for the analysis of the EN241170 fireball, which exhibited a long shallow trajectory. Given the image quality from 1970 and the specific characteristics of the fireball's trajectory, maximizing the effort in calibration was essential to obtain precise and reliable trajectory data.

We have defined 10 characteristic parameters for every camera station. These parameters are:
- Optical direction, az and alt; this is the direction which optical system center is pointing at.
- Optical center, x and y; this is the optical center in pixels in the scanned image and the origin is at the top left corner of the image.
- Tilt; this is the rotation of the camera around the optical axis.
- x/y-ratio; equals to 1 if an image is not stretched.
- f; focal length of the barrel-model.
- t; barrel-parameter of the barrel-model.
- $k_5$; correction parameter for the 5th degree of the barrel-model.
- $k_7$; correction parameter for the 7th degree of the barrel-model.

The parameters were fitted numerically using a minimal residual method. Correction terms $k_5$ and $k_7$ were fitted separately afterwards to avoid introducing undesired weight of correction terms and overfitting. The resulting parameters are shown in Table 3. During the fitting procedure, it became apparent that images were preserved in a good condition and did not suffer from severe stretching. Therefore, the x/y-ratio was excluded from being a free parameter and was set to 1 for every image. This is consistent with the observation of our Czech colleagues at the Ondřejov Observatory, who possess rich experience in digitizing photographic negatives and continue to use photographic films. They also employ semi-automatic measuring software for their measurements, which works in conjunction with scanned copies. To ensure the highest level of precision, our Czech colleagues conducted thorough tests on various scanners before their regular implementation in these measurements. Their findings indicate that the used in this work photogrammetric scanner Vexcel UltraScan 5000 consistently produces digital copies of metric quality comparable to direct measurements taken from the original negatives. Their tests also demonstrated that every other scanner, including high-resolution models, introduces distortions into the scanned copies (Spurný 2014).

Table 3. The derived camera parameters for the EN stations.

| EN station number | Stars used for calibrations | Optical direction, Az [deg] | Optical direction, Alt [deg] | Optical center, X [pix] | Optical center, Y [pix] | Tilt [deg] | f [pix] | t | k5 | k7 |
|---|---|---|---|---|---|---|---|---|---|---|
| 52 | 18 | 283.471 | 89.476 | 1996.331 | 1983.015 | -154.7963 | 1245.63 | 1.125 | -0.0089 | 0.0073 |
| 61 | 20 | 273.627 | 87.582 | 1970.508 | 2001.529 | -103.6618 | 1238.99 | 1.0969 | -0.0087 | 0.0081 |
| 62 | 18 | 267.030 | 89.628 | 1997.976 | 1990.582 | -147.0081 | 1255.604 | 1.1994 | -0.0055 | 0.0037 |
| 66 | 17 | 264.738 | 89.391 | 1989.431 | 1995.793 | -161.1053 | 1263.148 | 1.1913 | -0.0064 | 0.0048 |

6. **Retrieved atmospheric trajectory and orbit**

The reconstruction of the fireball trajectory was carried out as follows. The start point of the luminous flight was determined as early as possible, considering the background noise level in the images. However, due to limitations in the calibration process, the actual end points of the fireball were not directly measured, especially for low elevation angles where the accuracy was lower compared to higher elevation angles.

Potential inaccuracies that might arise due to the specific angle and position from which the fireball was observed were minimized by identifying the lowest point of the trajectory located at the same altitude as the lowest reference star utilized in the calibration process. Referred to here as the 'lowest referenced point', this point consistently proved to be closer to the actual terminal point of the fireball than to its beginning point. As a result, a significant part of the recorded trajectory was included in the triangulation process, enhancing accuracy. This method offers further advantages, especially in the context of the subsequent DFMC simulation, detailed in the following section. The DFMC simulation relies on an early start point within the luminous segment of the trajectory where the trajectory parameters can be determined with better precision, therefore the exact information about terminal height is not strictly required to generate the plausible strewn field (Moilanen et al. 2021). Consequently, our primary goal was to maximize the trajectory retrieval's precision by prioritizing the upper part of the trajectory over the very end of the luminous flight, which could potentially exhibit reduced accuracy below the lowest referenced points. The measured beginning and lowest referenced points are detailed in Table 4. These values were used with the plane intersection method (Ceplecha, 1987) to reconstruct the atmospheric trajectory. The resulting trajectory parameters are listed in Table 5.

During the analysis, it was observed that the fireball image from EN station 61 exhibited slightly sharper quality compared to images from the other stations. Moreover, its track in the sky had a favorable orientation for observation. Notably, the fireball trajectory did not intersect with the Moon trail, and its luminous flight was visible throughout its entire duration of 7.4 seconds. The image from EN station 61 was not affected by saturation and avoided being significantly blurred by atmospheric disturbances, particularly at low elevation angles. Considering these advantages and the overall image quality, the image captured by EN station 61 was chosen for measuring timed points along the fireball trajectory to derive the velocity during its luminous flight. The resulting velocity values are shown in Appendix B, along with a fitted exponential velocity curve (Whipple and Jacchia, 1957).

The measured velocity and height values obtained from the analysis of the image from EN station 61 were used to solve for the dimensionless ballistic coefficient α and mass loss parameter β values, determined based on the minimization of the sum of squares of the values on the left-hand side of equation (2) as described in (Gritsevich 2007, 2009, Peña-Asensio et al. 2023, Andrade et al. 2023):

$$2\alpha \, exp(-y) - \Delta \, exp(-\beta) = 0 \quad (2.1)$$

$$\Delta = \bar{E}i(\beta) - \bar{E}i(\beta v^2) \quad (2.2)$$

$$\bar{E}i(x) = \int_{-\infty}^{x} \frac{e^t dt}{t} \quad (2.3)$$

Equation 2 describes the relationship between the normalized height $y = h/h_0$ and the velocity $v = V/V_e$, where $h_0$ is the scale height of the homogeneous atmosphere with a value of 7160 m, and $V_e$ represents the initial entry velocity of the meteoroid at the start of its luminous trajectory (Stulov, 1997). An exponential integral, denoted as Ei(x), is a mathematical function defined as explained in the equation 2.3.

Due to the shallow trajectory slope, the resulting parameters α of 20.29 and β of 1.615 strongly suggest the likelihood of a meteorite fall after the EN241170 fireball event (Gritsevich et al. 2012, Turchak and Gritsevich 2014, Sansom et al. 2019a, Moreno-Ibáñez et al. 2020, Peña-Asensio et al. 2021, Boaca et al. 2022, Peña-Asensio et al. 2023). An inherent advantage of employing the α-β model, irrespective of the specific algorithm used, is the ability to determine the pre-atmospheric mass value and track how the mass of the main fragment changes along the trajectory by solely relying on meteor dynamics in the atmosphere (Gritsevich and Stulov 2006, Gritsevich 2008b, Lyytinen and Gritsevich 2016b). Incorporating photometry into the model, using the already known (determined based on dynamics) α and β values, allows for inferences regarding the values of the shape change coefficient μ and luminous coefficient τ (Gritsevich and Koschny 2011; Bouquet et al. 2014; Drolshagen et al. 2021a, 2021b). Since these values are not necessary for establishing the potential link between the EN241170 fireball event and the Ischgl meteorite, their determination is beyond the scope of the present study.

Table 4. Measured positions of the EN241170 fireball from the images. The lowest referenced point corresponds to the fireball's lowest trajectory point positioned at the same altitude as the lowest reference star point utilized during the calibration process.

| EN station number | beginning point direction, Az [deg] | beginning point direction, Alt [deg] | lowest referenced point direction, Az [deg] | lowest referenced point direction, Alt [deg] |
|---|---|---|---|---|
| 52 | 164.813 | 34.617 | 177.535 | 19.661 |
| 61 | 265.245 | 39.754 | 241.043 | 20.530 |
| 62 | 84.470 | 16.670 | 107.713 | 10.767 |
| 66 | 108.036 | 35.582 | 155.403 | 15.184 |

Table 5. The triangulated EN241170 fireball trajectory.

| Parameter | Value |
|---|---|
| Beginning latitude [deg] | 48.4345 |
| Beginning longitude [deg] | 11.2154 |
| Beginning height [km] | 77.646 |
| lowest referenced latitude [deg] | 47.5196 |
| lowest referenced longitude [deg] | 10.6008 |
| lowest referenced height [km] | 38.139 |
| Average heading [deg] | 204.443 |
| Average angle [deg] | 19.483 |
| Radiant, Dec [deg] | 55.2605 |
| Radiant, Rac [deg] | 237.3591 |

Table 6. The heliocentric orbit of the November 24, 1970 fireball was originally given by Ceplecha (1977) in epoch 1950.0. In our analysis, we are listing the parameters in epoch 2000.0.

| Parameters | Ceplecha 1977 (1950.0) | This work (2000.0) |
|---|---|---|
| a (AU) | 1.199 ±0.009 | 1.223 +/- 0.036 |
| e | 0.252 ±0.005 | 0.259 +/- 0.018 |
| q (AU) | 0.896 ±0.003 | 0.907 +/- 0.010 |
| ω (°) | 122.9 ±1.2 | 127.2 +/- 4.5 |
| Ω (°) | 241.071 | 241.7845 +/- 0.0001 |
| i (°) | 31.0 ±0.4 | 31.60 +/- 0.30 |
| M (°) | | 31.9 +/- 3.8 |
| perihelion | | 1970-10-11 T 00:06:12.445 |

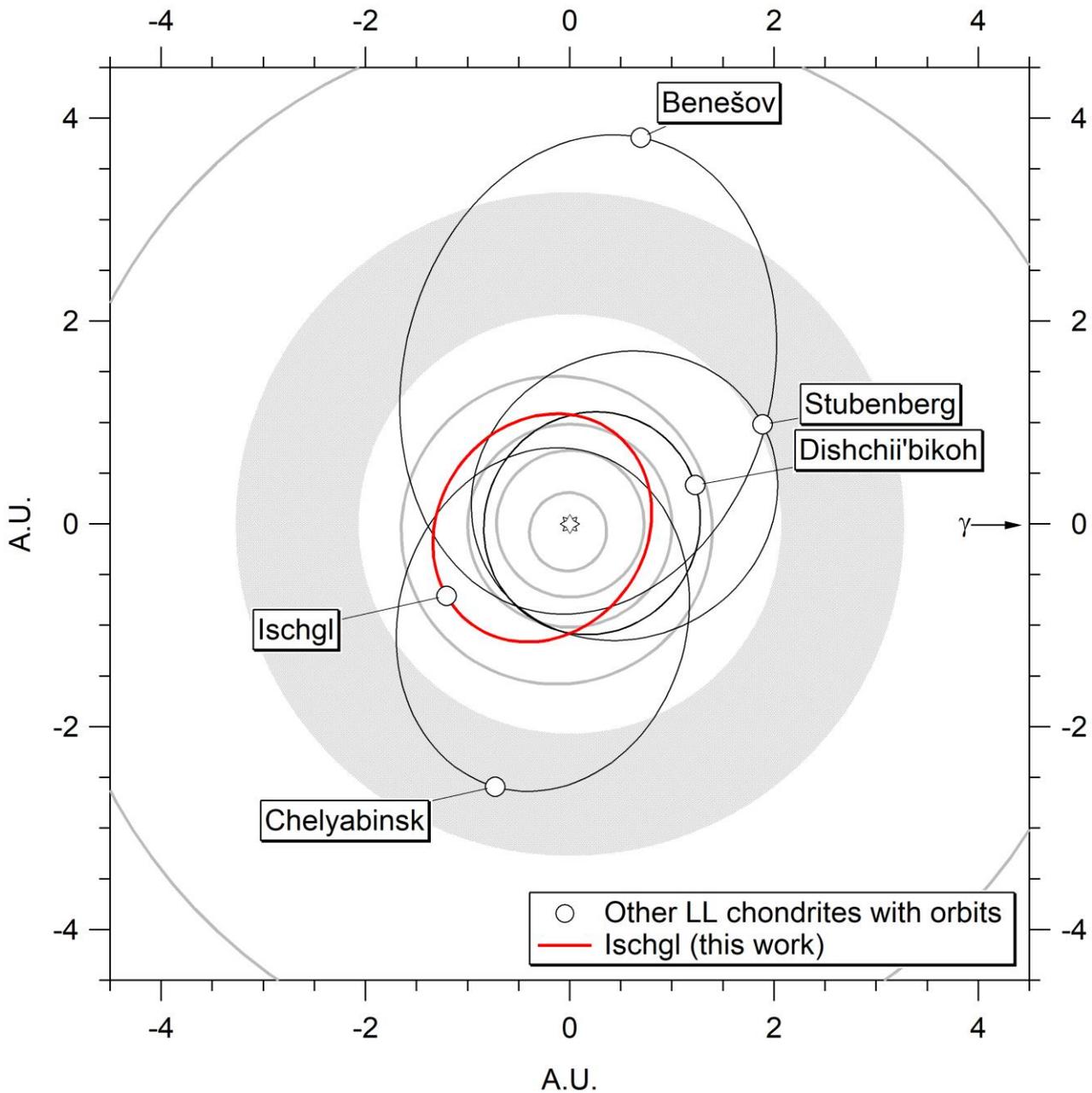

Figure 9: The calculated orbit of the Ischgl meteoroid (in red) and the four other LL ordinary chondrites with known orbits (black lines), together with the orbits of the five innermost planets (gray lines) and the asteroid belt (shaded area), projected onto the ecliptic plane from North. Each meteorite is labeled at the aphelion of its precursors orbit. The distinctive orbit of the Ischgl meteoroid sets it apart from the known orbits of the other LL ordinary chondrites, suggesting a different history for this specific case.

Considering values of the product of the pre-atmospheric shape factor multiplied by the drag coefficient $c_dA_e$ ranging from 1.5 to 1.8 (values commonly used to describe a more realistic complex-shaped body, see, e.g., Gritsevich 2008a, Trigo-Rodríguez et al. 2015, Gritsevich et al. 2017, Meier et al., 2017, Devillepoix et al. 2018, 2022a, Shober et al. 2022), as well as the measured bulk density value obtained from laboratory analysis and the shape change coefficient μ (needed for terminal mass estimate) of ~0.5 (Bouquet et al. 2014), the resulting α and β correspond to an object with a pre-atmospheric mass of 95-165 kg (corresponding to the characteristic radius of 19 to 23 cm), and a resulting terminal mass on the ground of 13.7-23.7 kg. A smaller value of the shape change coefficient μ (indicative of less fragment tumbling) or a smaller β value would imply less intense mass loss, leading to an increase in the terminal mass on the ground.

The EN stations were equipped with a rotating shutter operating at a frequency of 12.5 Hz (Oberst et al., 1998). To avoid unnecessary large deviations in the velocity distribution, we measured every second point during the time intervals of 0.16 seconds, with the start time set as 1970-11-24T01:47:00Z. The timed points with corresponding directions (azimuth, altitude) are listed in the Supplementary Material.

The heliocentric orbit was computed by integrating the meteor path while accounting for the gravitational influences of the Earth and the Moon. This calculation was performed using FireOwl, which gives similar orbital results when compared to both, already established Meteor Toolkit software and the 3D-FIRETOC used by the Spanish Meteor Network (Kyrylenko et al. 2023, Peña-Asensio et al. 2024). Our recalculation of the trajectory resulted in a heliocentric orbit similar to the one previously provided by Ceplecha (1977), as detailed in Table 6 and illustrated in Figure 9.

Given the significant advancements in numerical computations, modern approaches offer a more robust estimation of errors when determining pre-impact orbits of meteoroids compared to what could be achieved back in the 1970s (Dmitriev et al. 2015, Blanchard et al. 2022). In this study, as also in Kyrylenko et al. (2023), we adopted a method to assess uncertainties by calculating all possible combinations that result in the largest errors for the orbital solution. These errors are influenced by several factors, including the angular resolution of the camera, the calibration's angular resolution, residual of the calibration fit, the measured pixel coordinates of the fireball, and the velocity measurement errors. Quantitatively the results are comparable to a strict calculation of error propagation based on general rule of covariance transformation (Montenbruck and Gill 2000, Rice 2006, Dmitriev et al. 2015, Blanchard et al. 2022).

7. **The calculated strewn field**

To estimate the meteorite fall area on the ground, a Monte Carlo model is employed, which effectively combines the representation of processes occurring during the luminous trajectory and the dark flight phase (Moilanen et al. 2021). Monte Carlo simulations involve repetitive calculations using physical equations related to the phenomena. Plausible variations in parameters are generated using suitable probability distributions. In our model, the majority of variations are obtained using the normal distribution as a probability function, while some are generated using a modified version of the exponential distribution. By implementing these laws, the DFMC model effectively incorporates fragmentation events, and each generated fragment can be traced along its individual trajectory (Moilanen et al. 2021). Subsequently, DFMC models every fireball event as a suite of individual fragment trajectories (Moilanen and Gritsevich 2021a). This approach enables the determination of the entire probable strewn field through the utilization of importance sampling (Liu 2001, Särkkä 2013), in contrast to the earlier practice of repeatedly calculating the trajectory of a single fragment launched from the lowest observed fireball altitude.

The DFMC model incorporates considerations of gravity, air drag, the Earth's curvature, and atmospheric winds. The algorithm considers the mass constrained by the observed deceleration, ensuring that the model does not overestimate the total mass of the fragments on the ground. In many cases, this mass may be rightfully determined as zero. Every 0.04-second time step necessitates the recalculation of forces, spatial positions, ablation, and alterations in cross-sectional area caused by mass loss for each simulated fragment, as detailed in Moilanen et al. 2021. To produce realistic scenario in our simulations, Monte Carlo variations are introduced, as specified in Table 7. Fragment shape variations are achieved by utilizing different drag coefficient values, representing diverse drag forces associated with meteorite fragments of varying shapes.

The model has been earlier demonstrated through its application to historical examples of well-documented meteorite falls. These examples show a close match between the modeled strewn field and the actual distribution of the recovered meteorites, both in terms of fragment masses and their spatial dispersion on the ground. In particular, the model has been applied to meteorite falls such as the iron meteorite recovered in Sweden near the Ådalen village, Košice, Neuschwanstein, Winchcombe, and others, producing accurate results (Trigo-Rodríguez et al. 2015; Lyytinen and Gritsevich 2016b; Kohout et al. 2017; Maksimova et al. 2020; Jenniskens et al. 2021; Moilanen and Gritsevich 2021b; 2022; Kyrylenko et al. 2023). Furthermore, the model has played a critical role in supporting several successful meteorite recoveries, including Annama, Motopi Pan (asteroid 2018 LA), and Ozerki.

While photometry is a valuable tool for deducing a possible fragmentation history, accounting for meteoroid fragmentation can also be effectively addressed as described by Moilanen et al. (2021). Meteoroids undergo disintegration when the dynamic pressure surpasses the mechanical strength of the body (Silber et al., 2018; Tabetah and Melosh, 2018). Additionally, even a minor space-based impact can induce fractures, leading to the breakup of meteoroids during their subsequent atmospheric entry. This phenomenon is likely to occur already in low air pressure conditions, characterized by the Knudsen number (Moreno-Ibáñez et al., 2018).

Fragmentation introduces additional velocity components into the fragment trajectories, causing deviations from the original path (Barri, 2010). These velocity components, while relatively minor, are rapidly counteracted by atmospheric drag, allowing fragments to generally follow their initial trajectory. However, occasional noticeable downward deviations may occur when air drag decelerates smaller trailing fragments (Zhdan et al. 2005). The extent of alteration in direction and velocity is contingent on the fragment size, with larger fragments exhibiting lower separation velocities in the range of 10 to 190 m/s with an average value of 100 m/s (Moilanen et al. 2021). These separation direction and velocity changes are integrated with the fragment original velocity and direction to determine its subsequent path.

Following fragmentation, ablation reduces the size of the fragments, and some of them may completely ablate or vaporize. This process determines the expected mass and number of fragments reaching the ground. In the DFMC model, each simulated fragment undergoes some degree of fragmentation, accounting for the possibility of surviving fragments from earlier fragmentation events. To model fragmentation we employ a variation of the exponential distribution for a continuous random variable, which can be defined as:

$$f(\chi) = (1 - \mu)^i (1 - \chi) + \chi \qquad (3)$$

Here, $\mu$ determines the ratio of cumulative masses of the largest fragments produced to the original mass of the progenitor, and $\chi$ is a random test value between 0 and 1. The value of the exponent $i$ shapes the probability curve. Typical values for $\mu$ range between 0.02 and 0.05, while $i$ values usually fall between 4 and 7.

The mass of a newly produced fragment is determined by generating a random number between 0 and 1. If this number is less than or equal to the result of Equation 3, it becomes the normalized mass value used for the fragment after fragmentation. The new mass of the fragment is calculated by multiplying this random number by the mass of the progenitor. After determining the mass change for the fragment, its size is recalculated for use in the air drag equation. Fragments weighing less than 0.3 g are generally not significant in meteorite recovery (Gritsevich et al. 2014, Moilanen et al. 2021). While they may occasionally be found, it is often impractical to search for them. Therefore, in the DFMC code, fragments with mass below 0.3 grams are excluded from considerations.

While running the DFMC model, atmospheric conditions at the time of the event were taken into account, specifically air temperature, pressure, and wind velocity and direction from 25 km altitude to the ground. Fortunately, archive data from the German Weather Service (Deutscher Wetterdienst) for the night of November 24, 1970, specifically from the station of München–Oberschleissheim (DWD 10868) at 01h UT, were provided for our study by André Knöfel (see Appendix A). The wind direction was from the northwest to the north and was strongest at altitudes between 14–9 km, with speeds ranging between 20 m/s and 30 m/s. The wind direction data were missing for altitudes above about 20 km. Considering the thin air and the minimal effect of wind on a meteoroid of that size at those altitudes, we filled in the missing values by using the respective data from the nearest available altitude level.

Table 7. The DFMC input parameters for the November 24, 1970 fireball.

| Symbol | Parameter | Mean value | Error margins | Remarks |
|---|---|---|---|---|
| $\lambda_0$ | longitude | 10.931269°E | (see $e_0$) | WGS84 |
| $\varphi_0$ | latitude | 48.016017°N | (see $e_0$) | WGS84 |
| $h_0$ | altitude | 59 250 m | (see $e_0$) | |
| $e_0$ | spatial error of the start point | 0 m | ±300 m | |
| $δ_0$ | direction of trajectory | 204.49° | ±1.8° | 0° - 360° (0° = N, clockwise) |
| $γ_0$ | trajectory slope | 19.40° | ±1.8° | 0° - 90° (90° = vertical) |
| $V_0$ | velocity | 21 300 m s$^{-1}$ | ±300 m s$^{-1}$ | Velocity at the start point. |
| $a_0$ | deceleration limit | 111.8286 m s$^{-2}$ | - | Deceleration at the start point. |
| $h_{eh}$ | end height of fireball | 24 000 m | - | Ultimate cut-off value inferred from our analysis |
| $ρ_m$ | density of meteoroid | 3.31 g cm$^{-3}$ | ±0.5 g cm$^{-3}$ | Measured value for Ischgl used with our default error value. |
| $σ$ | ablation coefficient | 0.0018 s$^2$ km$^{-2}$ | - | after Moilanen et al. 2021 |
| $g_e$ | ground level | 2000 m | - | Altitude of the recovery site of the Ischgl meteorite. |

The parameters at the start point for our DFMC simulation of the November 24, 1970 fireball are presented in Table 7. By using a normal distribution as a probability function, the start point values for location and flight direction for each simulated fragment are calculated as described by Moilanen et al. (2021). Similar to other studied fireball cases, the simulation utilized a high start point along the luminous trajectory at an altitude of 59.25 km, which was determined approximately 2.56 seconds into the meteor stage, relying on the derived fireball trajectory and velocity as explained earlier in the paper. Strewn fields modeled from lower altitudes tend to miss earlier fragmentation events, resulting in narrow predictions that do not adequately constrain possible error margins in the fragments' trajectories due to fragment shapes and winds. Consequently, an 'idealized' strewn field prediction from a low altitude may match the actual meteorite strewn field in size and dimensions, but it might be geospatially located outside the real strewn field. To avoid this, we chose to initiate DFMC as early as possible on the trajectory, as soon as the deceleration value could be reliably established using the deceleration equation (16) in Gritsevich (2009). It is worth noting that sensible simplifications for this deceleration equation exist, obtained based on an approximation to the analytical solution of the meteor physics equations (Gritsevich et al. 2016), which may be useful for future reference.

The main concept of the DFMC model is the ability to produce a strewn field prediction covering an area where smaller meteorite fragments are likely to land, as they are usually more numerous, despite any hidden errors in the input wind and trajectory values. To achieve this, besides using a high start point, the DFMC default ±1.8° error margin for the trajectory direction and slope is employed (Moilanen et al. 2021). Present-day fireball trajectory calculations from videos and photographic records may be more precise; however, atmospheric data alone can be uncertain, leading to incongruence in strewn field models using different data sets. The default error margins in DFMC account for drifting caused by other forces due to fragment shapes, lift, specific motion of fragments (e.g., Magnus effect), and fragmentation events (Amin et al. 2019).

In this study, the primary focus was not exclusively on modeling the precise strewn field of the fireball, but rather on establishing a robust connection between the retrieved location of the Ischgl meteorite and the fireball event of November 24, 1970. Due to the absence of exact information about the fireball's terminal height, estimates were derived using the methodologies outlined in (Gritsevich and Popelenskaya, 2008; Moreno-Ibáñez et al. 2015, 2017; Gritsevich et al. 2016), and an ultimate cut-off altitude of 24 km was adopted to eliminate fragments that would not have completed their luminous flight by then. The remaining fragments produced a strewn field that also extended near the town of Ischgl, signifying that the Ischgl meteorite's recovery site is well within the leading part of the strewn field.

Although the utilization of a true digital elevation map (DEM) to create a more precise model of ground-level conditions would be optimal, especially for more recent meteorite falls, the primary objective of this study, which involves comparing the simulation results with the meteorite's 1976 recovery location, can be effectively achieved by employing a fixed ground level. Therefore, we have opted to conclude the free fall of all simulated fragments at an altitude of 2000 meters, approximately corresponding to the site where the Ischgl meteorite was recovered.

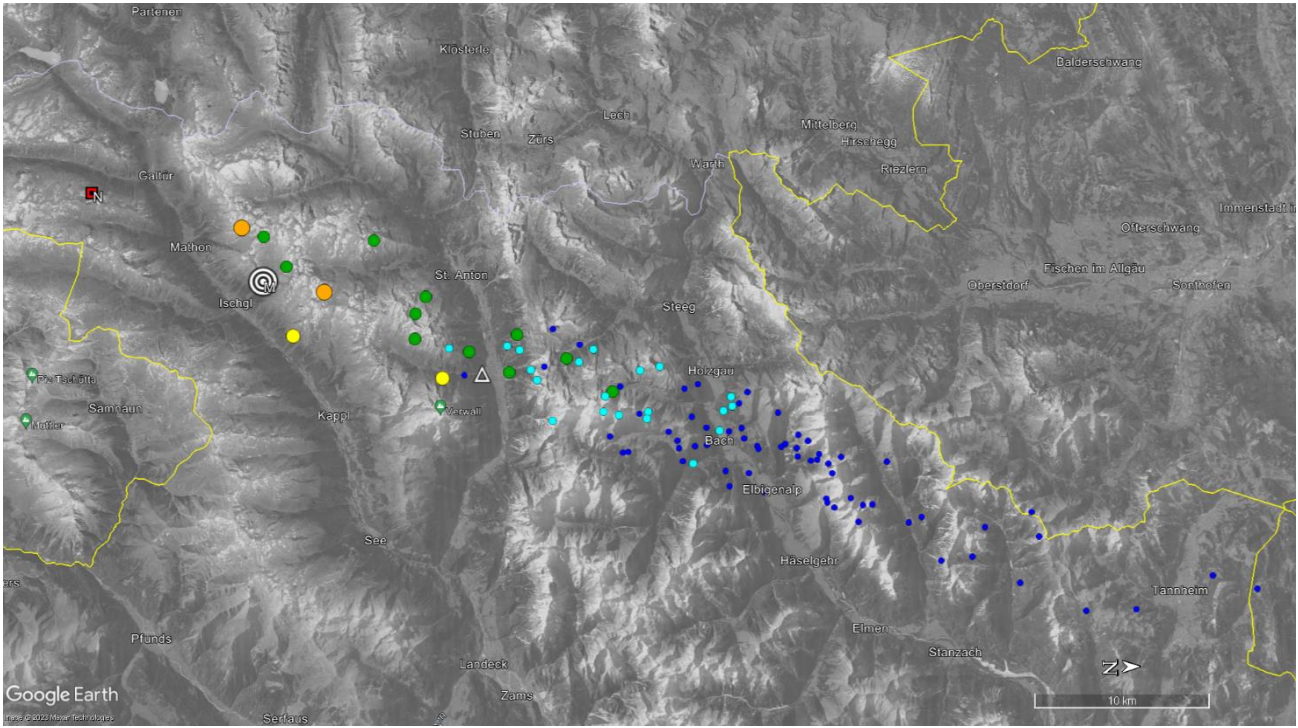

Figure 10. A DFMC simulation for the November 24, 1970 fireball (EN241170), using the parameters listed in Table 7. The location of the recovered Ischgl meteorite is represented by a white bull-eye symbol (M), while a predicted impact point by Ceplecha (1977) is marked as a white triangle. The red box (N) indicates the impact 'site' for the not fragmented nominal mass. Simulated fragments are color-coded as follows: orange for 3 – 10 kg, yellow for 1 – 3 kg, green for 0.3 – 1 kg, cyan for 0.1 – 0.3 kg, and blue for < 0.1 kg. The orientation is such that north is on the right.

In Figure 10, we present a DFMC simulation for the November 24, 1970 fireball, showing several larger fragments (0.472 – 5.106 kg) landing around the place where the Ischgl meteorite (M) was found in 1976. The probability of such a meteorite, unrelated to the fireball event, coincidentally landing in this specific area is extremely low, especially considering the date of the fireball and the recovery date of the meteorite. The strewn field extends over 65 km, and some of the smallest fragments might even have landed in Germany. However, from a practical standpoint, searching for fragments smaller than 100 g is not always worth the effort, as they are difficult to find unless there is a particularly easy-to-search ground in the area, such as a frozen surface of a lake. If we restrict the strewn field based on fragments larger than 100 grams, its size is approximately 7 km x 30 km.

In this specific simulation (Fig. 10), the main mass fragmented approximately 4 seconds after the start at an altitude of 30 to 40 km. Due to the shallow trajectory slope, this resulted in larger fragments separating more from the smaller ones. The nominal mass (N) trajectory was calculated as a reference, using the starting parameters from Table 7 and accounting for atmospheric data without the Monte Carlo variations (see Moilanen et al. 2021 for details). This 'reference location' with nominal mass (N) is purely theoretical, assuming mass loss conditioned to ablation in the absence of fragmentation.

8. **A heat map simulation**

The starting mass, or nominal mass, defined by the deceleration value at the altitude of 59.25 km corresponds to 92.4 kg. Utilizing the parameters listed in Table 7, we conducted a series of ten DFMC simulations. The results clearly demonstrate that the reported location of the Ischgl meteorite overlaps with the meteorite fall area produced by the EN 241170 fireball. The quantitative outcomes of these DFMC simulations can be summarized as follows:

1) When fragmentation occurs right after the start, the largest surviving fragment on the ground has an average mass of 3.06 kg, with a deviation ranging between 2.07 to 4.86 kg. The mean cumulative mass of DFMC simulated fragments on the ground is 26.07 kg, with a deviation of 8.50 to 37.61 kg.

2) If fragmentation of the nominal mass occurs 4-4.8 seconds after the start point, as shown in the example in Fig. 10, these values decrease. On average, the largest mass on the ground is 2.30 kg, and the cumulative mass of fragments on the ground is 24.30 kg.

Both scenarios align with the mass estimates obtained using the α and β approach as explained earlier in the text.

In Figure 11, we present the results of these ten DFMC simulations as a heat map illustration, a graphical representation that indicates the expected distribution of fragments' masses on the ground. For this simulation, the fragmentation events took place within the time frame of 4-4.8 seconds after the start, similarly to the earlier demonstrated single simulation run (Fig. 10). This intensified x10 simulation, conducted using the parameters outlined in Table 7, yielded a cumulative count of 3167 fragments. For a realistic representation of the results in Fig. 11, the obtained mass distribution of survived meteorite fragments was normalized by the number of simulation runs. Consequently, the heat map illustrates the average relative distribution of mass upon ground impact for a single simulation run.

These results serve as a valuable demonstration of how relying on rigid assumptions about atmospheric conditions and the behavior of an idealized individual fragment launched from the presumed "exact" terminal point of a fireball can limit our understanding of the size and shape of the actual strewn field. By concentrating solely on the calculation of a single-point impact location through precise dark flight trajectory calculations, one overlooks the broader perspective of the true extent of a meteorite fall area and the distribution of fragments on the ground.

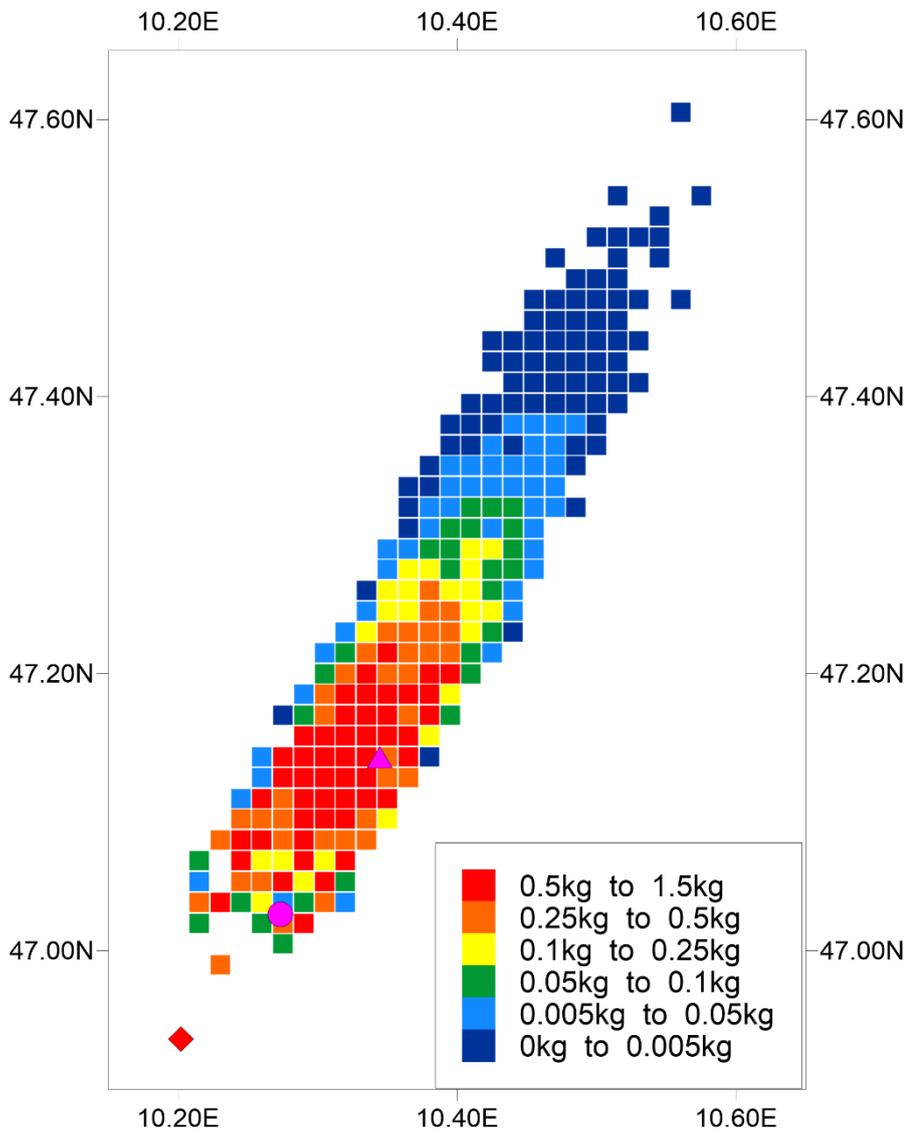

Figure 11. Enhanced DFMC simulation results are graphically presented as a heat map, depicting the anticipated dispersion of meteorite mass upon impact. Within the map, a rhombus symbolizes the nominal mass (same as N in Fig. 10) and a circle marks the recovered position of the Ischgl meteorite. Additionally, a triangle denotes the single point impact coordinates previously predicted by Zdeněk Ceplecha in 1977.

9. **Analysis of cosmogenic radionuclides and noble gases**

Cosmogenic nuclides can be used to derive the radius of the meteoroid which delivered a meteorite to Earth, the cosmic-ray exposure (CRE) age, i.e., the time the meteoroid was exposed to cosmic rays (mostly high-energy protons, colliding with the atoms of the meteorite material, inducing spallation and neutron-capture reactions, leading to the production of so-called "cosmogenic" nuclides (see, e.g., Herzog and Caffee, 2014)), as well as the time the meteorite spent on the Earth's surface. As detailed below, we measured two cosmogenic radionuclides ($^{60}$Co with a half-life of 5.2711 years, and $^{26}$Al with a half-life of 0.717 Ma) by decay counting in the whole available Ischgl meteorite sample, as well as all isotopes of the noble gases He, Ne, and Ar in two small (ca. 100 mg) sub-samples. Cobalt-60 is produced by neutron capture on $^{59}$Co nuclei in meteoroids large enough (a few 10 cm in radius and more) to develop a significant internal flux of secondary neutrons. After reaching the Earth's surface, the meteorite's cosmogenic radionuclides decay away over a few half-lives. The short half-life of $^{60}$Co means we can, at least if the parent meteoroid was large enough, measure the terrestrial residence time of the Ischgl meteorite, and thus test its association to the EN241170 (Mount Riffler) fireball. Unfortunately, however, more than eight half-lives of $^{60}$Co have passed between the fireball and the analysis of the $^{60}$Co activity of Ischgl in the lab. Assuming Ischgl derives from the Mount Riffler fireball, we expect only about 0.4% of the original $^{60}$Co activity to remain. On the other hand, the much longer half-life of $^{26}$Al (0.717 Ma) means its activity should still be essentially the same as on the day of the fall. If the meteoroid was exposed for at least a few Ma to cosmic rays in interplanetary space, its $^{26}$Al activity has reached

"equilibrium" (i.e., the decay rate matches the production rate) and can be used in combination with the concentration of the stable cosmogenic nuclide $^{21}$Ne to determine a nearly shielding-independent cosmic-ray exposure age ($^{26}$Al/$^{21}$Ne-method; e.g., Dalcher et al., 2013; Povinec et al., 2020). The remaining He, Ne, and Ar isotopes provide additional information on primordial noble gas components, the gas-loss history, and radiogenic gas-retention ages ($^{4}$He from the decay of U and Th and $^{40}$Ar from the decay of $^{40}$K).

The radionuclide analysis was carried out in the Low-Level Gamma-Ray Spectrometry Laboratory of the Department of Nuclear Physics and Biophysics of the Comenius University in Bratislava (Slovakia). A coaxial low-background High-purity Germanium (HPGe) detector (PGT, USA) with relative detection efficiency of 70% (for 1332.5 keV gamma-rays of $^{60}$Co) was used, while a NaI(Tl) detector was used for coincidence measurements. The detectors operated in a large low-level background shield with outer dimensions of 2 × 1.5 × 1.5 m (Povinec et al., 2009). To further decrease the detector background, an additional shield made of electrolytic copper (12×20×30 cm$^3$) was inserted inside the large shield. A contribution of the hard component of cosmic rays (muons) to the HPGe detector background has been partially eliminated by a plastic scintillation detector (50×50×5 cm$^3$), which operated above the small copper shield in anticoincidence with the analyzing HPGe detector (Kováčik et al., 2012). Special attention was given to the calibration of the detector efficiency, which may significantly influence the accuracy of radionuclide activities in odd-shaped meteorite samples. A Monte Carlo method, based on the GEANT 4 software package developed at the European Organization for Nuclear Research (CERN, see Agostinelli et al., 2003) was used. The radionuclide activities were calculated from the count rates in the full-energy peaks, corrected for background noise and self-absorption (Kováčik et al., 2013). Corrections for coincidence summing effects were applied for positron emitters ($^{26}$Al) and for the cascade transitions in $^{60}$Co. The uncertainties of the results are mainly due to counting statistics. The measuring time of the sample and the background was over 2 months. Several gamma-lines were identified in the gamma-ray spectrum of the Ischgl meteorite. We specifically looked for cosmogenic $^{60}$Co (gamma-energies at 1173.24 and 1332.50 keV), and for $^{26}$Al (annihilation peak at 511 keV and characteristic peak at 1808.65 keV). Results of the analysis are given in Table 8.

The noble gas analyses of two subsamples of Ischgl (Ischgl-1, 93.1 mg and Ischgl-2, 117.6 mg; Both samples from NHMW_#7946-F1c, i.e., a 1.09 g subsample from the main mass NHMW-N9269) were carried out at ETH Zurich in July 2014. Both sub-samples were wrapped in Al foil and loaded into the sample chamber of a custom-built, single-collector noble gas mass spectrometer (named "Albatros"; see Meier et al., 2017 for a detailed description of the instrument and analysis protocol). Extraction was performed in a Mo-crucible heated by electron bombardment. All isotopes of He, Ne, and Ar, as well as potentially interfering species (e.g., H$_2^{16}$O$^+$ for monitoring the interfering H$_2^{18}$O$^+$ on the signal of $^{20}$Ne$^+$ at mass 20 amu/e), were then measured in peak-jumping mode. Samples were bracketed by blanks (empty Al foils dropped into the crucible) and the whole run was started and ended with the measurement of known amounts of calibration gas. The results are shown in Table 8.

Table 8: Measured cosmogenic radionuclide activities in the bulk meteorite and noble gas (He, Ne, Ar) concentrations (including cosmogenic and radiogenic contributions to specific isotopes) measured in two sub-samples of the Ischgl meteorite.

| Nuclide or ratio | Half-life | Activity in bulk sample (dpm/kg) | Ischgl-1, 93.1 mg | Ischgl-2, 117.6 mg |
|---|---|---|---|---|
| Radionuclides | | | | |
| $^{26}$Al | 0.717 Ma | 52.8±3.0 | - | - |
| $^{60}$Co | 5.2711 a | 2.82±1.5 | - | - |
| Noble gases | | | | |
| $^{3}$He/$^{4}$He | - | | 0.0164±0.0001 | 0.0159±0.0001 |

| $^4$He (cm$^3$ STP/g) | - | 843±1 | 854±1 |
|---|---|---|---|
| $^{20}$Ne/$^{22}$Ne | - | 0.842±0.002 | 0.843±0.003 |
| $^{21}$Ne/$^{22}$Ne | - | 0.910±0.001 | 0.911±0.002 |
| $^{22}$Ne (cm$^3$ STP/g) | - | 3.221±0.004 | 3.183±0.004 |
| $^{36}$Ar (cm$^3$ STP/g) | - | 0.827±0.014 | 0.645±0.008 |
| $^{38}$Ar/$^{36}$Ar | - | 0.517±0.010 | 0.625±0.008 |
| $^{40}$Ar/$^{36}$Ar | - | 4310±70 | 4890±60 |
| $^3$He$_{cos}$ (cm$^3$ STP/g) | - | 13.87±0.12 | 13.54±0.05 |
| $^{21}$Ne$_{cos}$ (cm$^3$ STP/g) | - | 2.929±0.267 | 2.900±0.265 |
| $^{38}$Ar$_{cos}$ (cm$^3$ STP/g) | - | 0.310±0.003 | 0.321±0.002 |
| $^3$He/$^{21}$Ne$_{cos}$ | - | 4.73 | 4.67 |
| $^{22}$Ne$_{cos}$/$^{21}$Ne$_{cos}$ | - | 1.097 | 1.095 |
| $^4$He$_{rad}$ (cm$^3$ STP/g) | - | 769 | 782 |
| $^{40}$Ar$_{rad}$ (cm$^3$ STP/g) | - | 3567 | 3152 |

The $^{26}$Al activity of Ischgl is in the range typical for ordinary chondrites (Povinec et al., 2015a,b). Indeed, as we show further below, the noble-gas-based cosmic-ray exposure ages are ca. 7-9 Ma, suggesting that the activity of $^{26}$Al must have reached equilibrium and is thus representative of the production rate the meteorite sample has seen in the last few Ma before its fall.

The prevailing scenario suggests that as larger meteoroids enter the Earth's atmosphere, they undergo fragmentation, leading to individual fragments ablating and further fragmenting (Gritsevich et al. 2014; Jenniskens et al. 2022). In cases where complete ablation does not occur, this process leads to the production of meteorite fragments (Moilanen et al. 2021). These fragments, to some degree, exhibit a random sampling of the interior composition of their parent meteoroid. This stochastic sampling is consistent with the probability of meteorites predominantly originating from the outer layers of their progenitor, which harbor the majority of the mass. Using the files provided by Leya and Masarik (2009), we model the integrated activity of ca. 700 g (corresponding to a meteorite radius M = 3.7 cm) meteorites idealized as spheres and placed at variable depths (D) within a suite of meteoroids (with radii R between 10 cm and 150 cm), resulting in three classes of possible matches (Fig. 12): at shallow depths (D ~ M) in meteoroids with R = ca. 20 cm, at shallow depths in meteoroids with R = 85-150 cm, and at larger depths in meteoroids with R > 100-120 cm. The match with the R = 20 cm model is in excellent agreement with the sphere-equivalent radius derived from the fireball analysis (ca. 19-23 cm, from a mass of 95-165 kg and a density of 3.31 g/cm$^3$).

The $^{60}$Co activity of Ischgl (2.82±1.5 dpm/kg) measured in 2013 is too close to the detection limit (at ca. 1.2 dpm/kg) to be considered a reliable detection. The negligible $^{60}$Co activity confirms that Ischgl is a fall with a terrestrial age measured in decades at least, or that its parent meteoroid was small (R < 50 cm; a R = 20 cm meteoroid will have a $^{60}$Co activity of only about 0.5 dpm/kg for D ~ M (I. Leya, pers. comm.), which is already below the detection limit mentioned above). Although no strong case can be built on the non-detection of $^{60}$Co, the result is not inconsistent with Ischgl being related to the Mt. Riffler fireball.

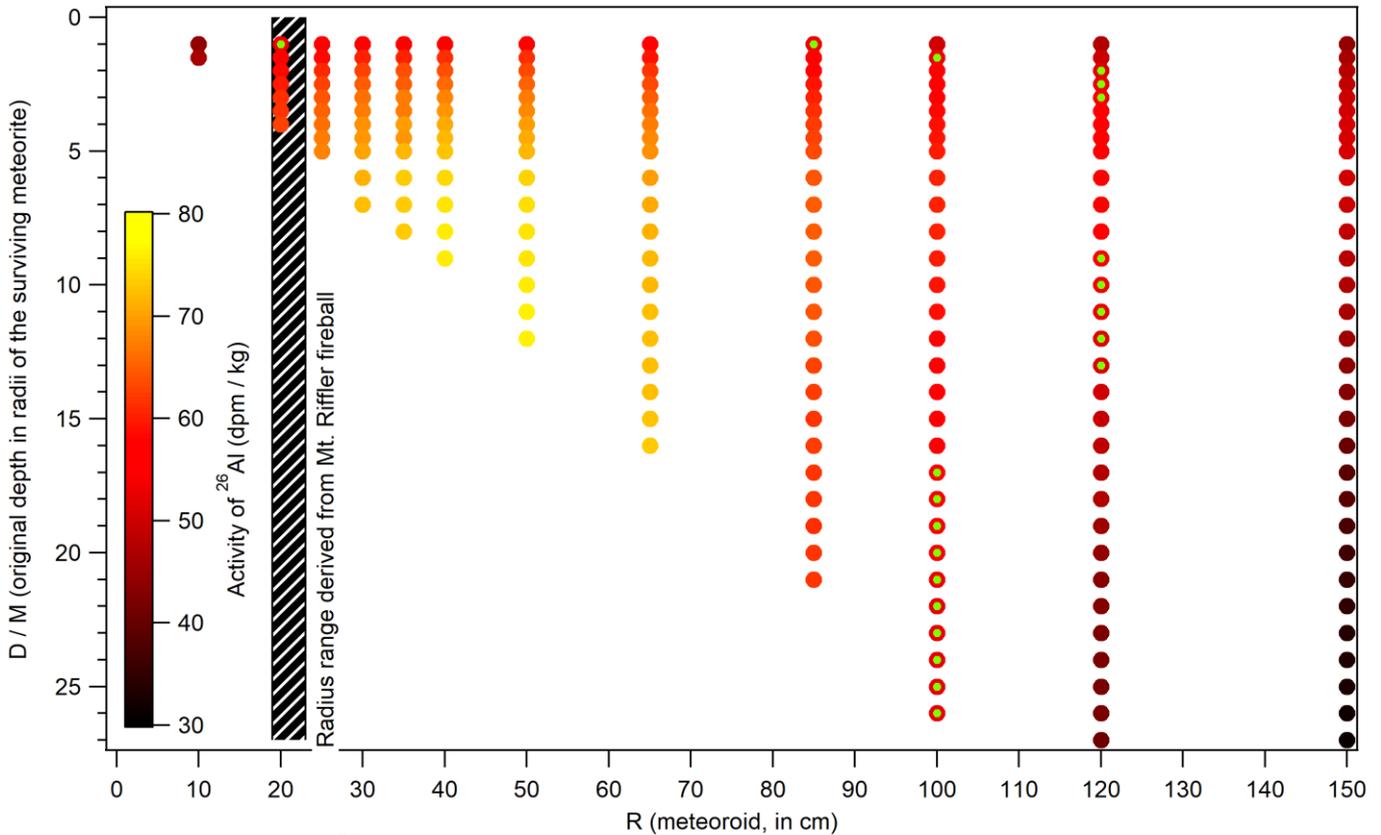

Figure 12. The activity of $^{26}$Al (shown on a color scale) as a function of meteoroid radius (horizontal axis) and depth of the sample, presented in multiples of the radius of the surviving meteorite (3.7 cm; on the vertical axis). The green dots represent radius-depth combinations that match the measured activity of 52.8±3.0 dpm/kg. The production rates shown are based on the model by Leya and Masarik (2009). The hashed area represents the most likely pre-atmospheric radius range of the meteoroid that caused the EN241170 (Mt. Riffler) fireball, as determined in this study.

The noble gas (He, Ne, Ar) inventory of Ischgl is typical of an equilibrated ordinary chondrite: almost completely cosmogenic and radiogenic, with only a minor contribution from trapped gases (most likely, of phase Q origin, e.g., Busemann et al., 2000), which is visible only in Ar. Therefore, all measured $^3$He is assumed to be cosmogenic, and all non-cosmogenic $^4$He (with $^4$He$_{cos}$/$^3$He$_{cos}$ = 5.3, based on $^{22}$Ne$_{cos}$/$^{21}$Ne$_{cos}$ = 1.096 and formula 6 in Leya and Masarik, 2009) is assumed to be radiogenic. For the determination of the cosmogenic $^{21}$Ne inventory, phase Q, air (Earth atmosphere) and a cosmogenic component with $^{22}$Ne$_{cos}$/$^{21}$Ne$_{cos}$ = 1.096 were adopted as endmembers of a three-component deconvolution. For Ar, a two-component deconvolution between air or phase Q ($^{36}$Ar/$^{38}$Ar = 5.32) and cosmogenic ($^{36}$Ar/$^{38}$Ar = 0.65) endmembers was used to determine the concentration of cosmogenic $^{38}$Ar. The measured $^{40}$Ar/$^{36}$Ar ratio in Ischgl is a factor of ~15 higher than in air, indicating that the majority of the measured $^{40}$Ar is from the in-situ decay of $^{40}$K and not from (adsorbed) terrestrial air. However, a small air contribution cannot be excluded: if all the non-cosmogenic Ar in Ischgl is from air (instead of phase Q), about 4-5% of the measured $^{40}$Ar would also be from air and not (in situ) radiogenic. However, since the determination of nominal K-Ar ages is wrought with additional uncertainties (e.g., the unknown K concentration, for which we adopted a typical value, see below), we assume in the following that all $^{40}$Ar is radiogenic. The determination of CRE ages is detailed in the next paragraph. To determine the radiogenic gas-retention ages, we need concentrations for U, Th and K. Although Brandstätter et al. (2013) presented the bulk chemical composition of Ischgl in 34 elements, only upper limits are given for the concentrations of Th and K in two samples, while U was not measured. Therefore, we adopt typical LL-chondritic U, Th, and K concentrations, of 13 ppb, 43 ppb and 790 ppm, respectively (Wasson and Kallemeyn, 1988), and determine the time needed to accumulate the measured concentrations of radiogenic $^4$He (from $^{235}$U, t$_{1/2}$ = 704 Ma; $^{238}$U, t$_{1/2}$ = 4.47 Ga; $^{232}$Th, t$_{1/2}$ = 14.1 Ga) and $^{40}$Ar (from $^{40}$K, t$_{1/2}$ = 1.25 Ga), respectively. Note that, as always, these "ages" assume complete retention of the decay products and have no meaning if radiogenic gas was partially lost through diffusion or shock at some point. We determine a $^4$He-retention-age of 2.3 Ga for both NG samples, while the $^{40}$Ar-retention ages are 3.4 and 3.6 Ga for samples Ischgl-1 and Ischgl-2, respectively. These Ga-scale retention ages are consistent (Stöffler et al., 1991) with the ones observed in meteorites with shock stage S3 (such as Ischgl).

The match of the measured activity with the R = 20 cm model is best for a depth of D / M ~1, i.e., the center of the eventual meteorite was at a depth of about one meteorite radius (3.7 cm) from the pre-atmospheric surface, which implies one side of the surviving meteorite must have been very close to that surface. Therefore, the two sub-samples in which noble gases were measured might be derived from any depth between 0 and 7.4 cm below the pre-atmospheric surface in an R = 20 cm meteoroid. Consequently, based on the LL-chondrite model by Leya and Masarik (2009), we expect production rates (in units of $10^{-8}$ cm$^3$STP g$^{-1}$ Ma$^{-1}$) for $^3$He, $^{21}$Ne, and $^{38}$Ar to be within ranges 1.56-1.80, 0.228-0.315, and 0.0334-0.0408, respectively. Combined with the measured concentration of these cosmogenic nuclides, these production rate ranges result in single-stage cosmic-ray exposure ages of 7.5-8.9 Ma, 9.2-13.0 Ma, and 7.6-9.8 Ma, respectively. While the nominal age ranges are not overlapping, when a typical uncertainty of the production rates on the order of 15% is included (Leya and Masarik, 2009) a CRE age of roughly 9 Ma would be consistent with the data. However, the cosmogenic $^{22}$Ne/$^{21}$Ne ratio measured in the two sub-samples, at 1.096, is too low to fit the shielding conditions described above (<7.4 cm depth in an R = 20 cm meteoroid), under which we would expect this ratio, which is often used as a shielding parameter due to its strong sensitivity to variations in shielding conditions, to be ca. 1.12 – 1.18 for LL chondrites. A lower $^{22}$Ne/$^{21}$Ne ratio suggests, in general, higher shielding, thus the observed difference is consistent with a multi-stage scenario, where the Ischgl meteorite was first exposed to cosmic rays in a somewhat larger meteoroid, which was then disrupted by a collision, and only a fragment of the original meteoroid reached the Earth (as the Mt. Riffler fireball), a few Ma after that collision, in order to allow the $^{26}$Al activity to enter into a new equilibrium consistent with the final size, as observed.

### 10. Weathering grade

Despite their ancient origin dating back to the early Solar System, meteorites are susceptible to alterations from the terrestrial environment with the Ischgl find being one of the least weathered examples. Weathering grade in the context of meteorites refers to a classification system that describes the degree of alteration a specimen undergoes from the time it enters the atmosphere to when it is analyzed in the laboratory. This grade reflects the extent of influences caused by chemical (water, air, salts), physical (temperature fluctuations, wind), and potentially biological (microbial activity) factors (Al-Kathiri et al. 2005).

Meteorites exhibit infinite variability in the time it takes for their recovery after the fall. Some meteorites are never recovered, while others may be found many years after their fall. The recovery timeline depends on many circumstances, such as the meteorite location, the likelihood of human witnesses, size and appearance of the sample, number of survived fragments, and the level of awareness about meteorite falls in the area where it lands. Meteorites falling in remote, inaccessible, or sparsely populated areas often remain undiscovered, particularly if the fall was not witnessed, well analyzed, or documented. Conversely, meteorites falling in deserts, where it is relatively easy to spot the sample, or densely populated regions are often recovered more promptly.

To characterize the degree of alteration in meteorite samples, several qualitative weathering indices have been introduced. One such approach involves using $^{57}$Fe Mössbauer spectroscopy to analyze the relative abundance of Fe$^{3+}$ in a sample, as unweathered ordinary chondrite falls typically contain negligible amount of ferric iron Fe$^{3+}$ (Bland et al. 1998, 2006). The most widely recognized weathering scale today is based on observations in polished thin sections of chondritic meteorites (Wlotzka 1993, Wlotzka et al. 1995, Al-Kathiri et al. 2005). This scale recognizes six degrees of weathering:

- W0: No visible oxidation of metal or troilite, but limonitic staining may be noticeable in transmitted light.
- W1: Small oxide rims around metal and troilite, with small oxide veins.
- W2: Moderate oxidation of metal (about 20-60% replaced).

- W3: Heavy oxidation of metal and troilite (60-95% replaced).
- W4: Complete (>95%) oxidation of metal and troilite, but no alteration of silicates.
- W5: Beginning alteration of mafic silicates, mainly along cracks.
- W6: Massive replacement of silicates by clay minerals and oxides.

Stony meteorites exhibit a higher susceptibility to weathering compared to iron ones. A prompt recovery after their fall or landing in conditions that minimize weathering may result in a W0 classification. Therefore, the Ischgl meteorite, given its W0 weathering grade, merits a distinct investigation, adding to the intriguing dimension of our study. Table 9 lists all Austrian meteorites and those found in relative proximity to Ischgl with a reported weathering grade in the Meteoritical Bulletin. The table has only a few entries, as the weathering grade for older falls was not determined at the time of the recovery.

The duration a meteorite retains a weathering grade W0 on the Earth's surface before recovery can vary largely. In some cases, meteorites with a W0 classification have been recovered quickly, while in others, there are extended periods between the fall and discovery. The Elbert LL6 case is an example where a fireball observed on January 11, 1998 resulted in the discovery of a W0 fragment over two years later in Colorado, USA. Noktat Addagmar LL5 ordinary chondrite was found in Mauritania in 2006, with the absence of short-lived radionuclides indicating a fall several decades prior. The Indian Butte H5 meteorite, recovered in 2013 from a fireball observed in 1998 by using weather radar signatures of historic falls, demonstrated that fragments from a 15-year-old fall may retain a W0 weathering grade. Finally, meteorites can exhibit remarkable preservation even over extended geological timescales. An illustrative example is the Morokweng LL6 meteorite, which, despite being ~144 million years old, retained its W0 classification. This preservation occurred through burial inside impact melt, demonstrating the diverse conditions under which meteorites can endure and maintain their original characteristics.

As of the beginning of March 2024, among the 174 records in the Meteoritical Bulletin containing the text "W0" (some of which are W0/1), 101 correspond to meteorite falls. Other interesting cases indicating a significant gap (of many years) between the meteorite fall and the recovery of a W0 sample include Souslovo and Sanggendalai. Such examples highlight the diverse circumstances surrounding meteorite falls, leading to the resilience of some specimens residing in certain terrestrial conditions to weathering. Al-Kathiri et al. (2005) suggested that meteorites weather more intensely in wet and humid areas, and older meteorites are generally more strongly weathered. Temperature fluctuations, particularly between summer and winter and day and night, may further affect the intensity of weathering. In high-altitude regions, such as the High Alps, where Ischgl is located, dry conditions prevail due to thin, cold air that holds minimal water vapor. The low temperatures also impede evaporation, contributing to an overall arid environment. Furthermore, historical records indicate a substantial snow depth (even below-average conditions entail more than 50cm of snow on the ground) at the Ischgl find location site (see Supplementary Material), with additional months of snowfalls occurring after November 24, effectively keeping the meteorite well within the snowpack.

The High Alps region, including the mountain above the Ischgl find location, consistently retains snow at elevations over 2000 m (Matiu et al. 2021). In satellite images taken during spring or summer seasons, persistent patches of snow can be seen on mountain slopes. This suggests that even these days and even in warmer seasons, certain areas maintain a snow cover, contributing to the unique environmental conditions of the region. Changes in temperature, precipitation patterns, and overall climate dynamics over the years influence the extent and persistence of snow cover. Therefore, the presence of year-round snow cover in the 1970s, especially at higher elevations, was probably more significant due to different climatic conditions.

| Name | Neuschwanstein I | Mürtschenstock | Kindberg | Ischgl |
|---|---|---|---|---|
| Linked to a fireball? | 6.4.2002 | N/A (find) | 19.11.2020 | 24.11.1970* |
| Month found | Jul-02 | Jul-17 | Jul-21 | Jun-76 |
| Exposure to temperatures > 0° and water | yes | unlikely too? | yes | unlikely |
| Type | EL6 | L6 | L6 | LL6 |
| Weathering grade | W0/1 | W1 | W1 | W0 |
| Mass** | 1.75 kg | 0.36 kg | 0.23 kg | ~1 kg |
| Coordinates | 47.52392° N, 10.80803° E | 47°4.285'N, 9°9.023'E | 47° 31.291' N, 15° 26.247' E | 47.02633° N, 10.27333° E |
| Elevation | 1650 m | 2000-2020 m | 920-940 m | 2000 m |
| Place | Bayern, Germany | Glarus, Switzerland | Steiermark, Austria | Tirol, Austria |
| Circumstances of recovery and fusion crust extent | A single black stone 10 x 9 x 6.5 cm, completely covered by fusion crust was found lying on the bare ground. No visible impact pit or hole was observed. The find location is 68.7 km northeast of Ischgl. Additionally, two more fragments were recovered, one approximately 1 km north (in an impact hole) and the other about 1.5 km southeast of the original site. For further details, refer to Oberst et al. 2004 and Moilanen et al. 2021. | Discovered during a mountain hike, a single black stone measuring 10.1 × 5.7 × 5.0 cm was found. It is nearly entirely covered by a black matte fusion crust. Small exposed areas, approximately 1 cm square, reveal an interior material of yellowish-brown color. The find location is 85.5 km west of Ischgl. | The fragment 8.1 × 5.2 × 3.2 cm, exhibits a dark brownish fusion crust covering approximately one-third of its surface. Discovered on the side of a private forest trail, the broken surface reveals a network of thin, dark shock veins that intersect the light greyish interior of the meteorite. The presence of orange-brownish oxide rims surrounding metal grains suggests a low weathering stage. The find location is 394.1 km east of Ischgl site. | A single black stone, completely covered by fusion crust, likely descended from a higher altitude. It was discovered by a forest ranger in the middle of a mountain road while clearing the remnants of a snow avalanche. |

Table 9. Relevant records of meteorites recovered in Austria, Germany, and Switzerland in geographical proximity from Ischgl, with known weathering
Additional notes: *This work. **Mass of the specimen at the time of recovery.

The duration a meteorite retains a weathering grade W0 may be influenced by various already established and potentially newly considered factors discussed below:

1. Quick recovery: Rapid retrieval minimizes exposure to terrestrial conditions, enhancing the chances of retaining W0 conditions. Shorter exposures are more conducive to preserving a sample. Meteorites that are recovered shortly after they fall are more likely to retain their original characteristics, including their fusion crust and volatile elements.
2. Exposure to water: Water exposure, such as rainfall or groundwater, accelerates weathering processes, with meteorites in arid regions more likely to be preserved due to minimal water sources. Wet samples are more likely to develop cracks due to their increased volume (Al-Kathiri et al. 2005).
3. Environmental conditions: The location where a meteorite lands can significantly affect its weathering grade. Certain environments slow down weathering processes, acting as protective factors. For example, in a desert, the dry climate and low moisture levels reduce the likelihood of chemical alteration. In polar regions, meteorites are preserved in ice, providing protection from weathering agents.
4. Geological conditions: Geological factors, such as the type of soil or rock in the impact area, can influence the preservation of a meteorite. Soils or sediments with high reactivity may cause more rapid alteration, whereas meteorites embedded in resistant geological formations may remain less weathered.
5. Local climate and seasonal variation: Regional weather changes influence meteorite weathering, hence understanding local climate conditions is essential for assessing chances for preservation. In regions with severe winters, low temperatures can lead to the formation of ice on the surface of the ground. Meteorites landing in these areas may become covered or partially embedded in ice during the winter months.
6. Mechanical abrasion: Sheltered meteorites experience less surface wear. Mechanical abrasion, influenced by natural forces like wind, can affect surface conditions.
7. Mechanical strength and composition: Certain minerals resist weathering, impacting meteorite preservation. Iron meteorites, with high iron content, tend to be more resilient than stony meteorites. On the contrary, the extreme fragility of CI chondrites makes them highly susceptible to terrestrial weathering, and they do not survive on Earth's surface for long after they fall. Five CI chondrites have been observed to fall: Ivuna, Orgueil, Alais, Tonk, and Revelstoke. Only a few others have been collected in Antarctica.
8. Size and mass: Larger meteorites, with greater volume and mass, have more internal material shielded from weathering, increasing the likelihood of retaining W0 classification, in particular when examining a thin section. Largest meteorites also tend to make deeper impact craters, which can help shield them from surface environmental factors.
9. Fusion crust: A well-preserved fusion crust acts as a protective barrier, shielding the interior of the meteorite. Subsequently it indicates minimal terrestrial exposure of the interior; therefore, meteorites with substantial fusion crust are more likely to retain a W0 grade. The characteristics of the fusion crust can vary among meteorites, even within the same strewn field, and even on a single fragment, depending on its flight history.
10. Atmospheric trajectory and velocity: Entry characteristics, character of motion, as well as fragmentation character and history may affect thickness of the fusion crust. A steeper entry angle or higher velocity may subject meteor body to greater heating and mechanical stress. Ablation rate and the time it took since last fragmentation affect fusion crust formation.
11. Microbial activity: Microbial communities can influence weathering grade, especially in environments conducive to certain microorganisms.
12. Human activity and urbanization: Proximity to human activity introduces potential disturbances, and urbanized areas may lead to pollution and contamination, affecting weathering grade. Remote, pristine areas with minimal human activity increase the probability of preservation, while impact in an area with high human traffic and industrial activity, may be subject to contamination or damage, higher exposure to salts and industrial pollutants.

13. Storage and handling: Proper preservation techniques after retrieval of the sample prevent its alteration and do also play a role. This includes storing the meteorite in a controlled environment to prevent further alteration, protecting it from moisture, and minimizing physical handling that could lead to contamination or damage.

The interplay of these factors collectively contributes to the overall preservation of meteorites. This indicates that the meteorite, covered in a deep layer of snow (that can be considered dry outside of its melting period) in an arid high-altitude environment, and retaining its intact fusion crust, likely never experienced prolonged exposures to positive temperatures or water, as these conditions could alter its appearance.

**Discussion**

Operational since 1966 until spring 2022, the German EN cameras continuously surveilled the night sky, encompassing an area of around $4*10^5$ km$^2$ on the ground. Over the course of almost six decades, only twelve instances of meteorite-dropping fireballs with reported terminal mass estimates exceeding 0.5 kg were simultaneously recorded by at least two camera stations (see Table 1 and Fig. 6). Among these events, our study identified the prominent fireball EN241170 as the most promising candidate for our research objectives. The primary focus of our investigation thus was to re-calculate the trajectory of EN241170 and assess its potential outcome, including the resulting strewn field.

To initiate the analysis, six out of the ten negatives containing records of fireball EN241170 were digitized using a high-quality photogrammetric scanner, and four of these images were employed to derive the trajectory solution. Calibration and adjustments for image distortions and exposure times were vital in ensuring accurate trajectory reconstruction. These calibration efforts establish a reliable basis for analyzing the fireball data captured by the camera stations and contribute to the accuracy of the subsequent analysis, identification and interpretation of a meteorite-dropping event. The combination of the radial distortion correction and the inclusion of higher-degree correction terms represents an advancement in the calibration process, making it a crucial step in accurately analyzing and interpreting the fireball trajectory data captured by the camera stations.

After retrieving the trajectory, the selection of the appropriate start point for the DFMC simulation involves several factors. It requires identifying a location where the most precise trajectory parameters can be extracted from fireball observations. This location usually differs from the point of maximum fireball brightness due to potential image overexposure. Furthermore, it is important to position the start point not too close to the end of the luminous flight, to effectively capture the earlier fragment separations. Already during the luminous flight, a meteoroid trajectory might not adhere to a straight line, exhibiting substantial deviations over longer distances, varying from hundreds of meters to several kilometers (Sansom et al. 2019b, 2020, Devillepoix et al. 2022). Therefore, a realistic model should have the capacity to represent these potential variations.

Incorporating the actual atmospheric data together with probable errors in the data into the Monte Carlo model significantly enhances the certainty of trajectory calculations accounting for different scenarios and improves the overall reliability of the predicted strewn field. Beginning the modeling process at an earlier stage plays a crucial role in ensuring the simulation accuracy, as it allows to comprehensively reproduce a broader range of the phenomenon. In our case, opting for a higher starting altitude takes into account the possibility of earlier fragmentation and accommodates model variations, including fragment shapes. This provides a more complete picture of the entire strewn field, rather than just a part or cross section of it or a single point. This comprehensive approach ensures that the model takes into account the complexities and uncertainties of atmospheric conditions, fragmentation, sizes and shapes, and the subsequent trajectories of individual fragments, contributing to a more accurate estimation of the meteorite fall area on the ground.

The advantage of using importance sampling in Monte Carlo simulations is that it significantly improves computational efficiency by focusing on the most influential factors while reducing the computational burden associated with less impactful variables (e.g. Kastinen 2020, 2022). By assigning higher importance to critical parameters, the simulation can provide more accurate results with fewer computational resources. This approach optimizes the use of computing power and time, making the simulation faster and more efficient while maintaining precision, especially when dealing with complex or computationally intensive models like analyzing meteor trajectories (Weryk and Brown 2012, Gural 2012, Sansom et al. 2015, 2017, Sansom 2016, Vida et al. 2020, Moilanen et al. 2021, Devillepoix et al. 2022a).

The DFMC model demonstrates the application of Monte Carlo methods to optimize computational processes, particularly in modeling the dark flight parts of the modeled atmospheric trajectories. This phase can be computationally consuming, with aerodynamic forces, influenced by the Reynolds number, presenting a significant challenge (Vinnikov et al., 2016, Havrila et al. 2021, Towner et al. 2022). The aerodynamic forces depend on the fragment shape, orientation, and motion characteristics, making them unique for each fragment. In certain situations, less common phenomena may occur, such as the Magnus effect, which emerges when a rotating object encounters a lift force perpendicular to the direction of its motion.

The main implication of our investigation is that the recovery site of the Ischgl meteorite is well inside the strewn field resulting from the DFMC simulation for the EN241170 fireball (Figures 10 and 11). Furthermore, Ceplecha's computed impact location (Ceplecha, 1977) is also within the DFMC modelled strewn field, albeit displaced to a slightly earlier landing along the trajectory compared to the coordinates of the recovered meteorite. Figures 10 and 11 illustrate that a line connecting Ceplecha's impact point to the Ischgl meteorite's discovery location runs exactly parallel to the DFMC strewn field direction as determined in our study, making the probability of unrelated meteorite falls almost negligible. The 1977 prediction was based by considering only one of the possible meteorite fragments, and due to the initial constraints set in the calculation, it landed earlier along the trajectory, suggesting a possible landing on Mt. Riffler, the tallest mountain in the Verwall Alps — a mountain range in the Central Eastern Alps (Fig. 10). Such a location would have made any meteorite recovery difficult, which is why no systematic search was ever undertaken in that area. Had the Ischgl meteorite been known soon after its recovery in 1976, the situation might have changed. However, it remained unknown for 32 years, until the recovery of the Neuschwanstein meteorite drew the attention of the public.

The Ischgl meteorite was found approximately 2000 meters above sea level, and its original landing position may have been higher up in the mountain (still well within the calculated strewn field, as seen from the surrounding topography in Fig. 10), taking into account the possibility of displacement caused by a snow avalanche. The highest point on the mountain slope above the discovery location, from which the meteorite might have been carried by an avalanche, stands at roughly 2975 meters. This location is situated about 1.6 kilometers north (with a horizontally projected distance of around 1.35 kilometers) from the meteorite's retrieved coordinates.

To compare our results with the reported recovery location of the meteorite, the ground level in the DFMC simulations was set to an altitude of 2000 meters too. The actual ground elevation within the predicted strewn field varies from 900 to 3000 meters, presenting a challenging search environment in this high-mountain terrain. Having an all-inclusive predicted strewn field that accounts for desirable Monte Carlo variations (Table 7) helps identify areas with a higher level of ease for conducting field searches, such as relatively open spaces, roads and river valleys, which can be prioritized for exploration.

Our findings highlight the potential for reanalyzing old cases using state-of-the-art techniques when sufficient photographic or recording data with reference coordinates are available. Although not capturing the lower part of the fireball trajectory may pose challenges, it does not preclude the possibility of predicting a strewn field (Moilanen et al. 2021). Moreover, while it is feasible to estimate the ground impact location for a fragment with a fixed size and shape using the observed trajectory endpoint, this approach is suboptimal. In addition to the aforementioned factors, it is essential to recognize the lack of accurate observational data concerning behavior of meteorite fragments during the dark flight phase, except for potential weather radar data (Fries and Fries 2010, Jenniskens et al. 2012). This absence of data provides no foundation for model comparison or validation. The meteor velocity during the luminous flight is significantly higher, and the duration of this phase is relatively short, providing insufficient data for predicting the quantity of surviving meteorite fragments, as well as their specific characteristics and behavior during the dark flight (Devillepoix et al.

2022b). This phase is characterized by a much longer duration and involves a transition from supersonic to subsonic flow regimes, characterized by considerably lower speeds (Moilanen et al. 2021, Silber et al. 2023). The primary limiting factor, therefore, in producing a realistic model of the phenomenon does not hinge on requiring the observed trajectory endpoint but instead on the availability of accurate enough atmospheric data corresponding to the time and location of the event, and relying solely on data from a single weather station is generally insufficient.

The recovery of the Ischgl meteorite emphasizes the value of continuous monitoring systems such as fireball networks set up around the globe. These long-term monitoring efforts provide valuable data on the frequency and characteristics of meteorite dropping fireballs, allowing for better predictions of strewn fields and potential recovery sites. With precise timing information to determine velocities and favorable geometry between the observer and the fireball, even single-station observations may provide a plausible trajectory (Lyytinen and Gritsevich 2013).

With the exception of unusually low-penetrating fireballs, such as the iron meteorite fall in Ådalen (Sweden) on November 7th, 2020, which exhibited a lowest observed altitude of 11.28 km (Kyrylenko et al. 2023), marking the deepest atmospheric penetration ever recorded for a fireball, the prevailing assumption for most fireballs involves a chondritic composition. This assumption is based on the statistical distribution of known meteorites. Since the Ischgl meteorite is an LL6 chondrite, for our earlier simulations we have used a default bulk density of 3.3 ±0.5 g cm$^{-3}$ (Moilanen et al. 2021). This mean value is typical for ordinary chondrites (e.g., LL type: 3.21, L type: 3.35, H type: 3.40, EL/EH types: 3.64 g cm$^{-3}$; Britt and Consolmagno, 2004), consistent with the study by Wilkison and Robinson (2000) that demonstrates that the bulk density of LL6 chondrites is 3.29 ±0.17 g cm$^{-3}$ and is in a good agreement with the bulk density of the Ischgl meteorite of 3.31 g cm$^{-3}$ determined and used throughout in this study. Our lower default bulk density for all chondrites compared to what studies have reported (3.375 ±0.255 g cm$^{-3}$) is due to the assumption that a meteoroid entering into the Earth's atmosphere can have higher porosity than surviving fragments (Meier et al. 2017). Porosity in the form of voids and fractures inside a meteoroid body likely acts as a trigger for fragmentation events, as atmospheric pressure conditions change during its penetration into the atmosphere.

Could the discovery of the Ischgl meteorite have been solely attributed to data captured by fireball cameras? Oberst et al. (1998) observed that, on average, the German part of the European fireball network had only 3 hours of dark and clear sky conditions per day. This represents just one-eighth of the total time during which a meteorite fall could potentially occur. It is clear that meteorites fall both during the day and night, under cloudy skies or when the Moon is prominently visible in the sky, times when the German cameras were not active. The same applies to other cases, including the successful recoveries of meteorites in Neuschwanstein, Stubenberg, and Renchen. In these instances as well, it is likely that meteorites could have fallen outside the exposure window of the German EN cameras. Consequently, an interesting avenue for further studies involves calculating the probability of multiple meteorite falls occurring in close geographic and temporal proximity. This analysis could also consider deducible constraints related to meteorite characteristics, including pre-atmospheric mass and the expected distribution of mass among surviving fragments.

The validation of the link between fireball observations and subsequent meteorite finds is typically established through the results of radionuclide analyses. These analyses offer support for the claim that the observed meteorite falls and the found meteorites correlate. Furthermore, it has been earlier demonstrated that the pre-atmospheric sizes, determined through radionuclide analyses, align well with the pre-atmospheric sizes independently derived from the deceleration analysis of the fireball (Gritsevich 2008a; Meier et al. 2017; Gritsevich et al. 2017). The combination of orbits with pre-atmospheric sizes, cosmic-ray exposure, and radiogenic gas retention ages, often referred to as cosmic histories, is of great significance because they can be used to constrain the meteoroid birth region and evaluate models of meteoroid delivery.

This study has also utilized cosmogenic nuclide analysis to gain insights into the Ischgl meteorite's origins and history. By measuring $^{60}$Co and $^{26}$Al activities in the meteorite sample and also analyzing noble gases, we have inferred its parent meteoroid's size and cosmic-ray exposure age. The results as well suggest a link between the Ischgl meteorite and the Mount Riffler fireball. The data also point to a multi-stage scenario, where the meteorite was initially exposed to cosmic-rays within a larger meteoroid, subsequently underwent a collision and fragmentation process, and eventually arrived on Earth. Notably, the measurements align the pre-atmospheric radius estimates of the Ischgl meteoroid with the demonstrated deceleration of the fireball.

As of 2024, the Ischgl meteorite is one of only nine confirmed meteorites (Mauerkirchen, Mühlau, Minnichhof, Lanzenkirchen, Prambachkirchen, Ischgl, Ybbsitz, Neuschwanstein (no. 3), and Kindberg), all ordinary chondrites, to have been recovered on the territory of present-day Austria. Within the last 90 years, only four Austrian meteorite names have been documented, including the confirmed instrumentally registered Neuschwanstein and Kindberg meteorite falls (Spurný et al., 2003, Oberst et al., 2004, Ferrière et al., 2022). The Ischgl meteorite presents a notably fresh appearance, and our conclusion is that it landed on Alpine snow cover on November 24, 1970. In this frigid high-altitude environment, characterized by extended periods of sub-zero temperatures and frequent heavy snowfalls, the meteorite was well-preserved, being buried within a layer of snowpack, and retaining its intact fusion crust. Eventually, it was carried downhill by a snow avalanche, making the find by Mr. J. Pfefferle possible. There were no visual reports of other bright fireballs in this area during the years after the EN241170 event and prior to the recovery of the Ischgl meteorite (Brandstätter et al., 2013).

While it is extremely unlikely, it is not totally impossible that there could be two meteorite falls (EN241170 and the Ischgl meteorite) so close to each other both geographically and in time, and both yielding kilogram-sized terminal mass fragments. However, with the support of the derived trajectory and calculated strewn field map, as well as cosmogenic nuclide analysis, we confidently conclude that the link between EN241170 and the Ischgl meteorite is as reliably established as it is for most other records of meteorites with an estimated pre-impact Solar System orbit.

**Conclusions**
Based on the excellent spatial and temporal agreement between the EN241170 (Mt. Riffler) meteorite fall and the subsequent meteorite find, further supported by the presence of fresh fusion crust, weathering grade W0, and additional laboratory data, we confidently conclude that the Ischgl meteorite originated from the fireball event that occurred on November 24, 1970. The previously predicted impact point near Mount Riffler, as published by Z. Ceplecha in 1977, falls well within our calculated strewn field and aligns with the correct trajectory direction. This is due to the differences and simplifications in dark flight modeling and presentation of results used in the past versus our Monte Carlo approach, which accounts for multiple fragment masses, a variety of shapes, and uncertainties in atmospheric data.

The recovery of the Ischgl meteorite and its association with the EN241170 fireball, which would have otherwise remained unknown, highlights the long-term value of continuous monitoring efforts and the importance of preserving observational and atmospheric data. It emphasizes the need to maintain and expand observational networks and digitize/open-source their output in order to enhance present understanding of meteorite falls and associated physical processes. Furthermore, it is important to timely and systematically identify meteorite falls from the vast amount of observational data as well as conduct subsequent field searches within the robustly calculated strewn fields.

The find in an Alpine snow-covered environment suggests that meteorites can be preserved and protected from significant weathering effects in certain high-altitude terrains, which stresses the importance of considering diverse geological settings beyond apparent challenges when conducting searches for meteorites. These environmental conditions, characterized by low humidity, low temperatures, and persistent snow cover, may contribute to the preservation of meteorites in a well-protected state. Indeed, the Ischgl meteorite, and potential additional fragments from that fall, remained undiscovered for many decades, emphasizing the need for timely and dedicated efforts to locate and recover meteorites. Had the meteorite been collected and identified soon after its fall, it could have provided valuable scientific insights into the composition and origin of the meteoroid and could have assisted in validating the used back then modeling approaches. Therefore, this case serves as a reminder of the significance of prompt reporting and investigation of fireball events to maximize the scientific potential of meteorite recoveries.

Our study presents compelling evidence supporting the reclassification of the Ischgl meteorite from the "find" to the "fall"/"confirmed fall" category in the Meteoritical Bulletin database. Following the cases of Příbram and Lost City, Ischgl is now the third oldest meteorite fall where a meteorite sample was recovered, and a photographic pre-impact heliocentric orbit was determined. It is reasonable to suggest revisiting other notable fireball events using the knowledge and techniques available today, and to explore how they correlate with the discoveries of meteorites. By combining historical records and modern modeling approaches, we gain further insights into the origins and characteristics of meteorite falls.

**ACKNOWLEDGEMENTS**

This work received support from the Academy of Finland project no. 325806 (PlanetS), the Spanish Ministry of Science, Innovation and Universities project PID2020-11, Junta de Andalucía grant P20_010168, the Severo Ochoa grant CEX2021-001131-S funded by MCIN/AEI/10.13039/501100011033, and the Slovak Science and Grant Agency VEGA project No. 1/0487/23. We are grateful for the support provided through the special-order contract D/957/67295888 with the DLR. Our sincere appreciation goes to the Ondřejov Observatory and the establishment of an effective all-sky meteor observation program in Germany, which made this study possible. We extend our gratitude to the retired staff at the Max-Planck-Institut für Kernphysik, Heidelberg. Special thanks are also due to Dr. Hugo Fechtig, former director at the Max-Planck-Institut für Kernphysik, and Günther Hauth (1933–2000), the MPIK technician, who dedicated three decades to deploying and caring for the German EN stations. Our heartfelt thanks go to the volunteer operators of the EN cameras for their patience and dedication over the years. We are grateful to André Knöfel and Michael Matiu for providing historical weather data and useful discussions. We thank Emilio Fernández García for his proactive and practical support during the preparation of this paper.

**DATA AVAILABILITY**

The data underlying this study are included within the article and its accompanying supplementary information.

Jansen-Sturgeon, T., Sansom, E.K. and Bland, P.A. 2019. Comparing analytical and numerical approaches to meteoroid orbit determination using Hayabusa telemetry. *Meteoritics & Planetary Science* 54:2149-2162. https://doi.org/10.1111/maps.13376

Jenniskens, P., Fries, M. D., Yin, Q. Z., Zolensky, M., Krot, A. N., Sandford, S. A., … Worden, S. P. 2012. Radar-enabled recovery of the Sutter's Mill meteorite, a carbonaceous chondrite regolith breccia. Science, 338(6114), 1583–1587. https://doi.org/10.1126/science.1227163

Jenniskens, P., Gabadirwe, M., Yin, Q.-Z., Proyer, A., Moses, O., Kohout, T., Franchi, F., Gibson, R.L., Kowalski, R., Christensen, E.J., Gibbs, A.R., Heinze, A., Denneau, L., Farnocchia, D., Chodas, P.W., Gray, W., Micheli, M., Moskovitz, N., Onken, C.A., Wolf, C., Devillepoix, H.A.R., Ye, Q., Robertson, D.K., Brown, P., Lyytinen, E., Moilanen, J., Albers, J., Cooper, T., Assink, J., Evers, L., Lahtinen, P., Seitshiro, L., Laubenstein, M., Wantlo, N., Moleje, P., Maritinkole, J., Suhonen, H., Zolensky, M.E., Ashwal, L., Hiroi, T., Sears, D.W., Sehlke, A., Maturilli, A., Sanborn, M.E., Huyskens, M.H., Dey, S., Ziegler, K., Busemann, H., Riebe, M.E.I., Meier, M.M.M., Welten, K.C., Caffee, M.W., Zhou, Q., Li, Q.-L., Li, X.-H., Liu, Y., Tang, G.-Q., McLain, H.L., Dworkin, J.P., Glavin, D.P., Schmitt-Kopplin, P., Sabbah, H., Joblin, C., Granvik, M., Mosarwa, B., and Botepe, K. 2021. The impact and recovery of asteroid 2018 LA, *Meteoritics and Planetary Science* 56: 844-893. https://doi.org/10.1111/maps.13653

Jenniskens, P., Robertson, D., Goodrich, C. A., Shaddad, M. H., Kudoda, A., Fioretti, A. M., and Zolensky, M. E. 2022. Bolide fragmentation: What parts of asteroid 2008 TC3 survived to the ground? *Meteoritics & Planetary Science* 57:1641-1664, https://doi.org/10.1111/maps.13892

Kastinen, D. 2020. The use of particle distributions in Solar system small body dynamics. Monthly Notices of the Royal Astronomical Society, 492(2), 1566–1578. https://doi.org/10.1093/mnras/stz3432

Kastinen, D. 2022 From Meteors to Space Safety: Dynamical Models and Radar Measurements of Space Objects. Ph.D. Thesis, Swedish Institute of Space Physics, Kiruna, Sweden.

Kero, J., Campbell-Brown, M.D., Stober, G., Chau, J.L., Mathews, J.D., and Pellinen-Wannberg, A. 2019. Radar observations of meteors. In: Galina O. Ryabova; David J. Asher; Margaret D. Campbell-Brown (Ed.), Meteoroids: sources of meteors on earth and beyond (pp. 65-89). Cambridge University Press.

Kohout, T., Haloda, J., Halodová, P., Meier, M. M. M., Maden, C., Busemann, H., Laubenstein, M., Caffee, Marc. W., Welten, K. C., Hopp, J., Trieloff, M., Mahajan, R. R., Naik, S., Trigo-Rodriguez, J. M., Moyano-Cambero, C. E., Oshtrakh, M. I., Maksimova, A. A., Chukin, A. V., Semionkin, V. A., Karabanalov, M. S., Felner, I., Petrova, E. V., Brusnitsyna, E. V., Grokhovsky, V. I., Yakovlev, G. A., Gritsevich, M., Lyytinen, E., Moilanen, J., Kruglikov, N. A., and Ishchenko, A. V. 2017. Annama H chondrite—Mineralogy, physical properties, cosmic ray exposure, and parent body history. *Meteoritics & Planetary Science* 52:1525–1541. https://doi.org/10.1111/maps.12871

Kováčik A., Sýkora I., and Povinec P. P. 2013. Monte Carlo and experimental efficiency calibration of gamma-spectrometers for non-destructive analysis of large volume samples of irregular shapes. *Journal of Radioanalytical and Nuclear Chemistry* 298:665–672. https://doi.org/10.1007/s10967-013-2509-8

Kováčik A., Sýkora I., Povinec P. P., and Porubčan V. 2012. Non-destructive gamma-spectrometry analysis of cosmogenic radionuclides in fragments of the Košice meteorite. *Journal of Radioanalytical and Nuclear Chemistry* 293:339–345. https://doi.org/10.1007/s10967-012-1667-4

Kyrylenko I., Golubov O., Slyusarev I., Visuri J., Gritsevich M., Krugly Yu. N., Belskaya I., and Shevchenko V. G. 2023. The First Instrumentally Documented Fall of an Iron Meteorite: Orbit and Possible Origin. *The Astrophysical Journal* 953:20 (10 pp). https://doi.org/10.3847/1538-4357/acdc21

Leya I., and Masarik J. 2009. Cosmogenic nuclides in stony meteorites revisited. *Meteoritics & Planetary Science* 44:1061–1086. https://doi.org/10.1111/j.1945-5100.2009.tb00788.x

Liu J. S. 2001. Monte Carlo Strategies in Scientific Computing. Springer.

Lyytinen, E., and Gritsevich, M. 2013. A flexible fireball entry track calculation program. Proceedings of the International Meteor Conference 2012, edited by M. Gyssens and P. Roggemans (Hove: International Meteor Organization): 155-167.

Lyytinen, E., and Gritsevich, M. 2016a. Calibration of occasionally taken images using principles of perspective. Proceedings of the International Meteor Conference 2016, edited by A. Roggemans & P. Roggemans (Hove: International Meteor Organization):159-163.

Appendix A: Atmospheric Data

We received archive atmospheric conditions data from the German Weather Service (Deutscher Wetterdienst) for the night of November 24, 1970, specifically from the München–Oberschleissheim station (DWD 10868) at 01h UT, coinciding with the event. The provided data included air temperature, pressure, humidity, wind velocity, and wind direction, and was generously supplied by André Knöfel. During the time of the meteorite fall, the prevailing wind direction ranged from northwest to north, with the strongest speeds observed between altitudes of 14 to 9 km, measuring between 20 m/s and 30 m/s.

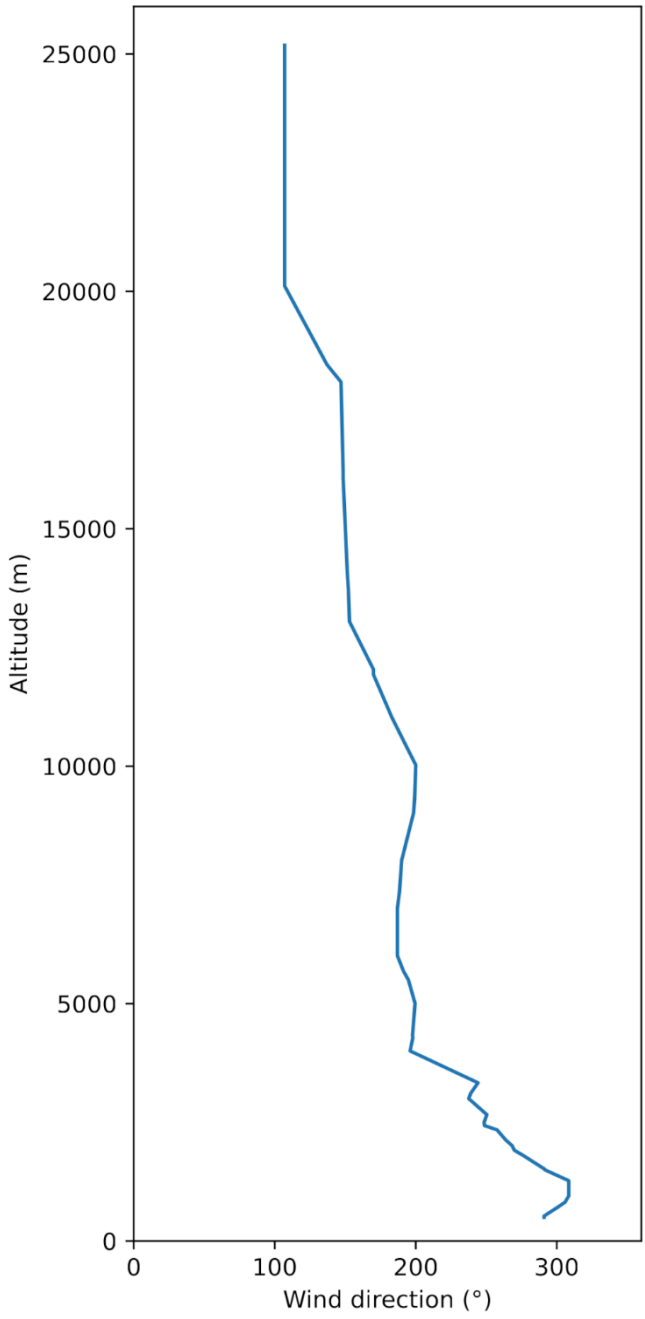 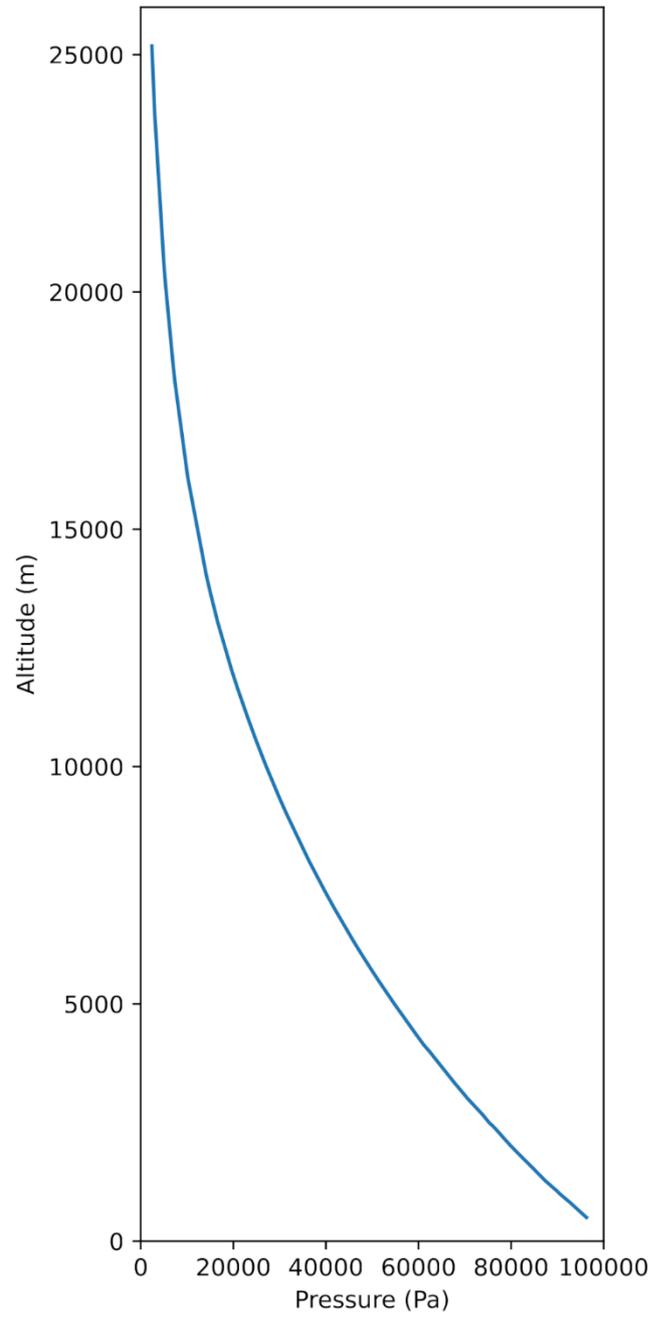

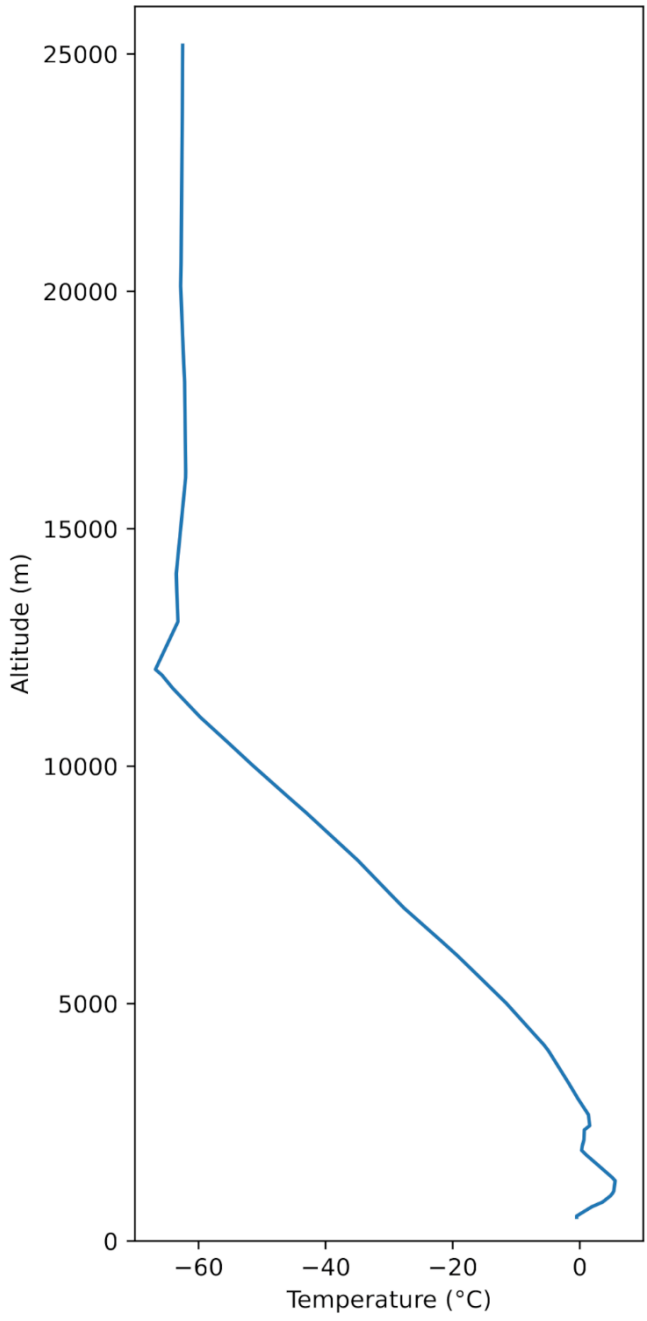

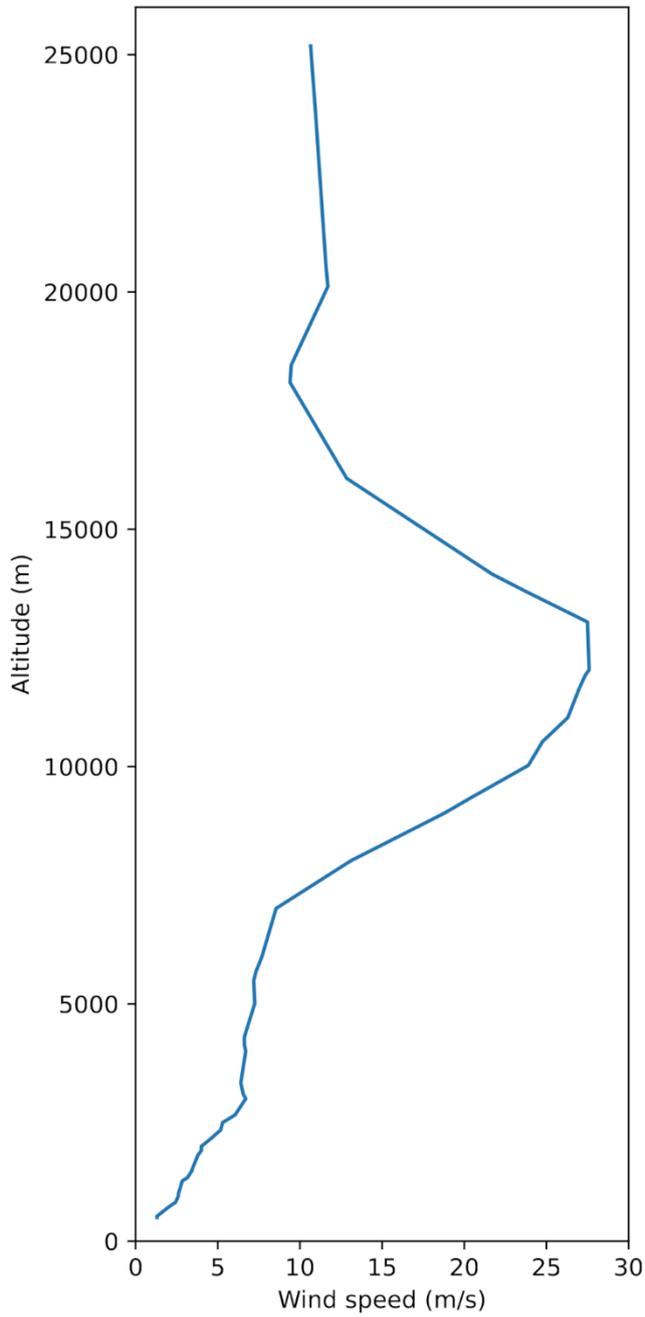

Figure A1. Wind direction (a), pressure (b), air temperature (c), and wind speed (d) relative to altitude data from the German Weather Service for the night of November 24, 1970 at 01:00 UT, sourced from the München–Oberschleissheim station (DWD 10868).

Appendix B: Velocity profile

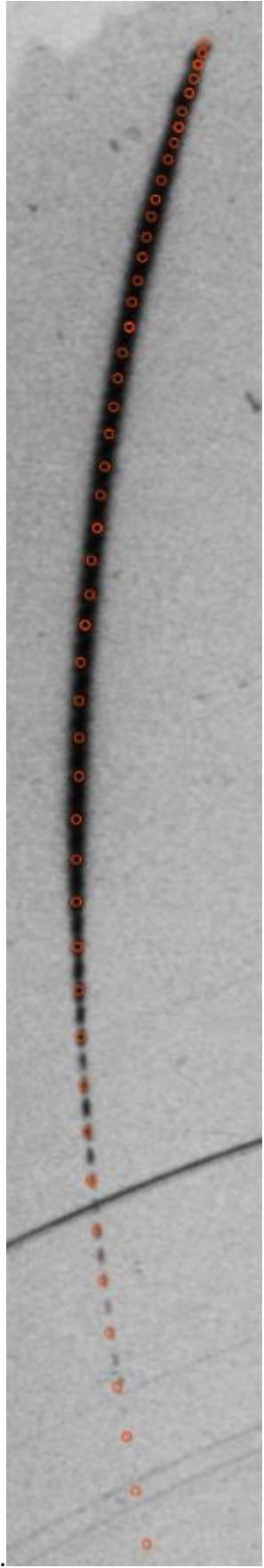
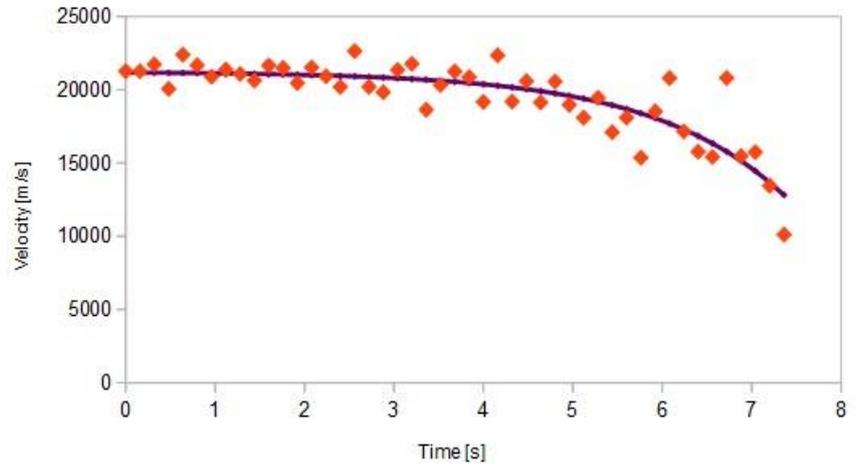

Figure A2. (a) The fireball, as recorded by EN station 61, with timed points along its trajectory. (b) The corresponding velocity values are plotted alongside a curve representing an exponential velocity fit, as proposed by Whipple and Jacchia (1957). The fitting process is based on the least squares method to derive optimal parameters for the fit, which were found as follows: b = 21245.05, c = -84.19, and k = 0.68.

48